\documentclass[12pt]{article}
\pdfoutput=1
\usepackage{makeidx}
\makeindex
\usepackage[a4paper]{geometry}
\usepackage{jheppub,amsmath,amssymb,amsfonts,amsxtra,mathrsfs,graphics,graphicx,amsthm,epsfig,ytableau,bm,longtable,float,color,tikz,mathtools,xfrac,footnote,rotating,lscape}
\usepackage{dsfont}
\usepackage{textgreek}
\usetikzlibrary{decorations.pathmorphing}
\usetikzlibrary{decorations.markings}
\usetikzlibrary{arrows, decorations.markings, calc, fadings, decorations.pathreplacing, patterns, decorations.pathmorphing, positioning}
\usepackage{tikz-cd}

\usepackage{fixmath} % for \mathbold
\usepackage{scalerel}
\newlength\bshft
\def\fakebold#1{\ThisStyle{\ooalign{$\SavedStyle#1$\cr%
  \kern-\bshft$\SavedStyle#1$\cr%
  \kern\bshft$\SavedStyle#1$}}}

\usetikzlibrary{positioning,shapes}
\usetikzlibrary{chains}
\usetikzlibrary{arrows,fit,decorations.pathreplacing}
\tikzstyle{every picture}+=[remember picture]
\tikzstyle{na} = [baseline=-.5ex]

\usepackage{empheq}
\usepackage{multirow}
\usepackage{booktabs}
\usepackage[american]{babel}

\usepackage[latin1]{inputenc}

\makeatletter
\newcommand{\vast}{\bBigg@{1}}
\newcommand{\Vast}{\bBigg@{5}}
\makeatother

\usepackage{hyperref}

\numberwithin{equation}{section}

\newcommand{\cf}{\textit{cf.}}
\newcommand{\eg}{\textit{e.g.}}

\numberwithin{equation}{section}

\newcommand{\be}{\begin{equation}} \newcommand{\ee}{\end{equation}}
\newcommand{\bea}{\begin{equation} \begin{aligned}} \newcommand{\eea}{\end{aligned} \end{equation}}

\def\U{\mathrm{U}}

\def\SU{\mathrm{SU}}

\newcommand{\rd}{\mathrm{d}}

\DeclareMathOperator{\Tr}{Tr}

\newcommand{\cI}{\mathcal{I}}

\newcommand{\cM}{\mathcal{M}}
\newcommand{\cN}{\mathcal{N}}

\newcommand{\cV}{\mathcal{V}}

\newcommand{\fg}{\mathfrak{g}}

\newcommand{\fn}{\mathfrak{n}}

\newcommand{\fR}{\mathfrak{R}}

\usepackage[Symbol]{upgreek}
\usepackage{bm}

\usepackage{calligra}
\DeclareMathAlphabet{\mathcalligra}{T1}{calligra}{m}{n}

\theoremstyle{plain}

  \theoremstyle{definition}

\providecommand{\examplename}{Example}
\providecommand{\theoremname}{Theorem}

\makeatletter
\g@addto@macro\bfseries{\boldmath}
\makeatother

\title{Proving the equivalence of $c$-extremization and its gravitational dual for all toric quivers}

\author[a]{Seyed Morteza Hosseini,}
\author[b,c]{and Alberto Zaffaroni}
\affiliation[a]{Kavli IPMU (WPI), UTIAS, The University of Tokyo, Kashiwa, Chiba 277-8583, Japan}
\affiliation[b]{Dipartimento di Fisica, Universit\`a di Milano-Bicocca, I-20126 Milano, Italy}
\affiliation[c]{INFN, sezione di Milano-Bicocca, I-20126 Milano, Italy}
\emailAdd{morteza.hosseini@ipmu.jp}
\emailAdd{alberto.zaffaroni@mib.infn.it}

\preprint{IPMU19-0006}

\abstract{The gravitational dual of $c$-extremization for a class of $(0,2)$ two-dimensional theories obtained by twisted compactifications of D3-brane gauge theories living at a toric Calabi-Yau three-fold has been recently proposed. The equivalence  of this construction with $c$-extremization has been checked in various examples and holds also off-shell. In this note we prove that such equivalence holds  for an arbitrary  toric Calabi-Yau. We do it by  generalizing the proof of the equivalence between $a$-maximization and volume minimization for four-dimensional toric quivers. By an explicit parameterization of the R-charges we map the trial right-moving central charge $c_r$ into the off-shell functional to be extremized in gravity. We also observe that the similar construction for M2-branes on $\mathbb{C}^4$ is equivalent  to the $\mathcal{I}$-extremization principle that leads to the microscopic counting for the entropy of magnetically charged black holes in AdS$_4\times S^7$. Also this equivalence holds off-shell.}

\begin{document}

\setcounter{tocdepth}{2}
\maketitle

%*******************************************************************************
%
%     Main body
%
%*******************************************************************************

\date{Dated: \today}

% \hypersetup{
% colorlinks,breaklinks,
%             linkcolor=black
% }

% \tableofcontents

% \hypersetup{
% colorlinks,breaklinks,
%             linkcolor=[rgb]{0,0,0.7}
% }

\section{Introduction}
\label{sect:intro}

Central charges play an important role in the study of superconformal field theories (SCFTs) in even dimensions. In supersymmetric gauge theories the R-symmetry current is not necessarily unique and mixes with the global symmetry currents. This  happens in particular in most models with a holographic dual. It is well known that, for ${\cal N}=1$ supersymmetric theories in four dimensions,  the extremization of a trial  central charge $a$
with respect to a varying R-symmetry allows to identify the  exact R-symmetry of the superconformal theory \cite{Intriligator:2003jj}.\footnote{The exact R-symmetry is the one appearing in the superconformal algebra.} Similarly, for $\cN = (0,2)$ supersymmetric theories in two dimensions, the extremization of a right-moving trial central charge $c_r$ allows to identify the exact R-symmetry \cite{Benini:2012cz,Benini:2013cda}.
The gravity dual of $a$-maximization is the volume minimization principle discovered in \cite{Martelli:2005tp,Martelli:2006yb}.\footnote{See also \cite{Tachikawa:2005tq,Szepietowski:2012tb} for a different approach based on five-dimensional supergravity. See also a similar approach for $c$-extremization in \cite{Karndumri:2013iqa}.} 
The equivalence of $a$-maximization and volume minimization has been proven in \cite{Butti:2005vn} for all quivers associated with D3-branes at toric Calabi-Yau three-fold singularities and generalized in \cite{Lee:2006ru,Eager:2010yu}. On the other hand, the gravity dual of $c$-extremization has been recently found in a series of very interesting papers \cite{Couzens:2018wnk,Gauntlett:2018dpc}. The authors of \cite{Couzens:2018wnk,Gauntlett:2018dpc} have checked the equivalence of their formalism with $c$-extremization
in various explicit examples. It is the purpose of this note to prove this equivalence for all theories obtained by twisted compactifications of D3-branes sitting at an arbitrary toric Calabi-Yau three-fold, just by generalizing   the arguments of \cite{Butti:2005vn}. 
 
The main focus in this note are theories that are obtained by a twisted compactification of four-dimensional ${\cal N}=1$ superconformal theories living on D3-branes sitting at the tip of 
a toric Calabi-Yau cone  $C(Y_5)$ over a Sasaki-Einstein manifold $Y_5$. These four-dimensional theories are well known and classified in terms of the toric data \cite{Hanany:2005ve,Franco:2005rj, Feng:2005gw}.
The gravitational dual is AdS$_5\times Y_5$. When compactified on a Riemann surface $\Sigma_\fg$ with a topological twist parameterized by magnetic fluxes $\fn_a$, the theory can flow in the infrared (IR) to a $\cN = (0,2)$ CFT.
The gravity solution dual to such  CFT is a  warped background AdS$_3\times_W Y_7$, where $Y_7$ is topologically a fibration of $Y_5$ over $\Sigma_\fg$, with a five-form flux. 

Given the close similarity between the gravitational dual of $a$- and $c$-extremization, let us start by first reviewing the story for  $a$-maximization in four-dimensions. 
By relaxing the equations of motion but still imposing the conditions for supersymmetry, the authors of \cite{Martelli:2005tp,Martelli:2006yb} defined an off-shell 
class of supersymmetric backgrounds obtained by replacing   $Y_5$ with a general Sasaki manifold. The background depends on a Reeb vector, $b=(b_1,b_2,b_3)$, which specifies the direction of the R-symmetry
inside the three isometries of $Y_5$. It has been shown in \cite{Martelli:2005tp,Martelli:2006yb} that the extremization of the volume of the Sasaki manifold identifies the exact R-symmetry of the CFT and allows to compute its central charge. The proof that this procedure is equivalent to $a$-maximization involves choosing  a convenient parameterization of the R-charges of the toric quiver in terms of the toric data and define a natural parameterization of the R-charges in terms of the Reeb vector \cite{Butti:2005vn}
\bea
 \label{4dparamet0}
 \Delta_a(b_i) = \frac{\pi \text{Vol} (S_a(b_i) )}{b_1 \text{Vol} (Y_5(b_i))} \, ,
\eea
where $S_a$ are toric three-cycles in $Y_5$. One then shows that
\bea \frac{\pi^3 N^2}{4 \text{Vol}(Y_5 (b_i))} 
 \label{a:toric:general0}
 \equiv a (\Delta_a) \Big |_{\Delta_a(b_i)} \, ,
\eea
thus demonstrating the equivalence of $a$-maximization and volume minimization.
Notice that the equivalence holds not only for the extremal value but is valid {\it off-shell}, since
the two expressions in \eqref{a:toric:general0} are equal for generic values of $b_i$.
There is an important difference between the two extremization principles. $a$-maximization is performed on the space of all R-symmetries.  This spans the three mesonic symmetries, associated with the isometries of $Y_5$ and a number (in principle large) of baryonic symmetries, associated with the non-trivial three-cycles of $Y_5$. On the other hand, volume minimization is performed  on the direction of the Reeb vector, spanned by $b_i$ and corresponding to the mesonic symmetries only. The consistency of the two extremizations is a consequence of the automatic decoupling of the baryonic symmetries from the $a$-maximization procedure in the given parameterization. This follows from the identity proved in \cite{Butti:2005vn}
\bea\label{id20}
\sum_a  B_a \frac{\partial a (\Delta_a)}{\partial \Delta_a}\Big |_{\Delta_a(b)} \equiv 0 \, ,\eea
where $B_a$ is a baryonic symmetry.  

After compactification on $\Sigma_\fg$ we obtain a two-dimensional theory depending on magnetic fluxes $\fn_a$ for all the symmetries of the original theory, including the baryonic ones. The exact R-symmetry can be found by extremizing the trial right-moving central charge with respect to the mesonic and baryonic symmetries \cite{Benini:2012cz,Benini:2013cda}. There is a simple formula for the trial central charge of the $(0,2)$ CFT at large $N$, that, in the basis for R-charges of \cite{Butti:2005vn}, reads  \cite{Hosseini:2016cyf} 
\bea
 \label{trial-c0}
 c_r(\Delta_a,\fn_a) & = -\frac{32}{9} \sum_{a=1}^d \fn_a \frac{\partial a(\Delta_a)}{\partial \Delta_a}  \, .
\eea
In order to study the gravitational dual of $c$-extremization, the authors of \cite{Couzens:2018wnk,Gauntlett:2018dpc} defined a family of off-shell backgrounds, again depending on the Reeb vector.
They also defined a functional $c(b_i,\fn_a)$ of the Reeb vector and fluxes whose  extremization selects the on-shell R-symmetry.  It has been explicitly checked in many examples in \cite{Couzens:2018wnk,Gauntlett:2018dpc} that this procedure is equivalent to $c$-extremization, and the equivalence holds off-shell.
We will prove in this note that this is true in general for all toric quivers and that the proof \cite{Butti:2005vn} extends very nicely to the two-dimensional case.
Indeed, the ingredients are exactly the same. We will define a natural parameterization of the R-charges in terms of the Reeb vector and magnetic fluxes, $\Delta_a(b_i,\fn_a)$,
just by generalizing the logic behind \eqref{4dparamet0}. Then we will show that for an arbitrary toric quiver
\bea
 \label{id10}
  c(b_i,\fn_a) \equiv c_{r} (\Delta_a, \fn_a) \Big |_{\Delta_a(b,\fn)} \equiv
 - \frac{ 32 }{9} \sum_{a=1}^d \fn_a \frac{\partial a (\Delta_a)}{\partial \Delta_a} \Big |_{\Delta_a(b,\fn)} \, .
\eea
Moreover, as in four dimensions, the baryonic symmetries explicitly decouple from the extremization process in this parameterization
\bea
 \label{id20}
 \sum_a  B_a \frac{\partial c_{r} (\Delta_a, \fn_a)}{\partial \Delta_a}\Big |_{\Delta_a(b,\fn)}  \equiv 0 \, .
\eea
 In particular, we do \emph{not} see any particular difference in the role of baryonic symmetries in two dimensions compared to four.
 
It is also interesting to study the theories living on M2-branes  at a toric Calabi-Yau four-fold $C(Y_7)$ and their twisted compactifications on a Riemann surface.
In this case, the exact R-symmetry of the three-dimensional theory is obtained by extremizing the free energy on $S^3$,  $F_{S^3}(\Delta_a)$.
The equivalence of volume minimization for four-folds \cite{Martelli:2005tp,Martelli:2006yb} and the extremization of $F_{S^3}(\Delta_a)$
has been checked in many examples in \cite{Herzog:2010hf,Jafferis:2011zi}. Given the complications of three dimensions and the absence of a complete classification of quiver duals to Calabi-Yau four-folds, there is no general proof. 
The twisted compactifications of M2-brane theories are dual in the IR to AdS$_2\times Y_9$ backgrounds, where $Y_9$ is topologically a fibration of $Y_7$ over $\Sigma_\fg$.
These backgrounds can be interpreted as the horizon of magnetically charged AdS$_4$ black holes. The construction in \cite{Couzens:2018wnk} also applies to these solutions
and the authors of \cite{Couzens:2018wnk}  identified the quantity to extremize with the entropy of the black hole in various cases. Interestingly, it is suggested by a field theory computation \cite{Hosseini:2016tor}
that the entropy of magnetically charged black holes in AdS$_4\times Y_7$  should be obtained by extremizing the functional 
\bea
 \label{iextr0}
 \cI(\Delta_a,\fn_a) = -\frac{1}{2}  \sum_{a=1}^{d} \fn_a \frac{\partial F_{S^3}(\Delta_a)}{\partial \Delta_a} \, .
\eea
This is certainly true for the theory with $Y_7=S^7$ as shown in \cite{Benini:2015eyy,Benini:2016rke}, where a microscopic counting for the entropy of magnetically charged black holes in AdS$_4\times S^7$ has been performed. We then expect that, also off-shell,  the construction of \cite{Couzens:2018wnk} is dual  to $\cI$-extremization. 
 In this note we just verify this statement for $Y_7=S^7$, reproducing the extremization of \cite{Benini:2015eyy,Benini:2016rke} also off-shell. We leave the investigation of more general Sasaki-Einstein manifold $Y_7$, where the computation is more complicated, to the future. The microscopic computation  of the entropy of  black holes in AdS$_4\times Y_7$ for generic $Y_7$ is still an open problem. In particular, baryonic symmetries enter in a puzzling way in the large $N$ limit, as noticed in \cite{Hosseini:2016tor,Hosseini:2016ume,Azzurli:2017kxo}.  The formalism of \cite{Couzens:2018wnk,Gauntlett:2018dpc} seems well suited to address these problems and we hope to come back to these questions in the future. Finally, notice the analogy of \eqref{iextr0} with \eqref{trial-c0}.  In the context of the large $N$ limit of topologically twisted theories these identities arise as special cases
of the index theorem discussed in \cite{Hosseini:2016tor,Hosseini:2016cyf}.

The note is organized as follows. In section \ref{CFT} we discuss general features of four-dimensional toric quivers and their twisted compactifications. In section \ref{sec:c} we first review the proof of the equivalence between $a$-maximization and volume minimization for all four-dimensional toric theories and  then we extend it to the equivalence between $c$-extremization and the construction in \cite{Couzens:2018wnk,Gauntlett:2018dpc}.
For the convenience of the reader, the technical aspects of the proof are deferred to appendix \ref{proof}.
In section \ref{sec:examples} we give explicit formulae for the R-charge parameterization and we present few examples.
In section \ref{sec:BH} we show that the formalism \cite{Couzens:2018wnk,Gauntlett:2018dpc} for $Y_7=S^7$ is equivalent off-shell to the $\cI$-extremization principle for black holes in AdS$_4\times S^7$.
Finally, in appendix \ref{c versus a} we review the proof of \eqref{trial-c0} for the right-moving central charge $c_r$. 

\section{Introducing the field theory}\label{CFT}

In this section we review some general aspects of the quiver gauge theories living on D3-branes at toric Calabi-Yau singularities and of their twisted compactifications
on Riemann surfaces. 

\subsection{${\cal N}=1$ superconformal field theories}\label{sec:4dSCFT}
We first discuss the four-dimensional aspect of the story. Consider  the type IIB background  AdS$_5\times Y_5$, where $Y_5$ is a five-dimensional Sasaki-Einstein manifold. In the AdS/CFT correspondence, this is dual to the ${\cal N}=1$ superconformal theory living on $N$ D3-branes sitting at the tip of the Calabi-Yau  cone CY$_3=C(Y_5)$ with base $Y_5$  \cite{Klebanov:1998hh,Acharya:1998db,Morrison:1998cs}. 
Familiar examples of Sasaki-Einstein manifolds include $T^{1,1}$, whose dual is the Klebanov-Witten theory  \cite{Klebanov:1998hh}, and the  Y$^{p,q}$ and L$^{p,q,r}$ spaces 
\cite{Gauntlett:2004yd,Gauntlett:2004hh,Cvetic:2005ft}, whose dual field theories have been identified in \cite{Benvenuti:2004dy} and \cite{Benvenuti:2005ja,Butti:2005sw,Franco:2005sm}, respectively.
When the CY$_3$ is toric, there is a general prescription for constructing the gauge theory associated with the D3-branes \cite{Hanany:2005ve,Franco:2005rj, Feng:2005gw} based on dimer models and tilings.
For our purposes, we will need just some general information about the quiver, that we review following \cite{Butti:2005vn}.

A toric affine CY$_3$ is specified by its fan, a collections of vectors $v_a$ in $\mathbb{R}^3$ with integer entries. The Calabi-Yau condition requires that all the $v_a$ lie on a plane that we will take to be
the plane orthogonal to the vector $e_1=(1,0,0)$.  The toric cone is then specified by $d$ vectors $v_a=(1,{\vec v}_a)$ for $a=1,\cdots\, , d$. The restriction to the plane of these vectors define  a regular polygon with integer vertices called the toric diagram. There is a toric divisor $D_a$ for each vertex.  Each $D_a$ is a cone over a three cycle $S_a$ in $Y_5$. There are $d$ such cycles but only $d-3$ are independent in cohomology.  All the data and symmetries of the gauge theory can be extracted from the geometry \cite{Hanany:2005ve,Franco:2005rj, Feng:2005gw}. The theory has an R-symmetry and $d-1$ $\U(1)$ global symmetries that can mix with it.  
A particularly important role is played by the baryonic symmetries. There are precisely $d-3$ of them, corresponding to the inequivalent non-trivial three-cycles $S_a$ of $Y_5$. They are holographically dual to
the $d-3$ gauge fields that we obtain by reducing the type IIB four-form potential on the three-cycles $S_a$. The remaining three symmetries are called mesonic and are holographically dual to the three gauge fields
associated with the isometries of the toric $Y_5$. One is an R-symmetry and the other two are global symmetries.  A convenient way to parameterize the global and R-symmetry comes from the prescription in \cite{Butti:2005vn} or, equivalently,
from the folded quiver formalism of \cite{Benvenuti:2005ja}. The $d-1$ global symmetries can be parameterized by assigning a real number $F_a$ to each vertex with the constraint
\bea
 \label{global}
 \sum_{a=1}^d F_a =0 \, .
\eea
In the minimal toric phase,%
\footnote{There are many different quivers that describe the same IR SCFT. They are related by Seiberg dualities.
The toric phases  have  the same number of gauge groups but different matter content.
The minimal phase corresponds to the quiver with the smallest number of chiral fields.}
the theory contains a number $|G|$ of gauge group factors $\SU(N)$ equal to twice the area of the toric diagram.
Moreover, defining the vectors $w_a=v_{a+1}-v_a$ lying in the plane, there are precisely $|(e_1,w_a,w_b)|$ bi-fundamental chiral fields $\Phi_{ab}$ with charge $F_{a+1}+F_{a+2}+\ldots+ F_{b}$
for each pair $(a,b)$  such that the outgoing normal of $w_a$ can be rotated counter-clockwise  into that of $w_b$ in the plane with an angle smaller than $\pi$.%
\footnote{In this note, we will use the following notation for determinants of  vectors
$(v_a,v_b,v_c) \equiv \det (v_a,v_b,v_c)$. We also  identify indices modulo $d$, so that, for example,  $v_{d+1}=v_1$.
%\langle {\vec v}_a, {\vec v}_b \rangle &\equiv  \det (e_1, v_a, v_b) \, ,
} 
%where $e_1=(1,0,0)$. 
The baryonic symmetries, which we will denote by $B_a$, are further characterized by the vector identity
\bea
 \label{baryonicsymm}
 \sum_{a=1}^d B_a v_a  =0 \, .
\eea
Similarly, we can parameterize the R-charges of all fields in the quiver by assigning a number $\Delta_a$
to each vertex with the constraint \cite{Butti:2005vn}
\bea
 \label{sumDelta}
 \sum_{a=1}^d \Delta_a =2\, .
\eea 
The chiral fields $\Phi_{ab}$ have R-charge $\Delta_{a+1}+\Delta_{a+2}+\ldots +\Delta_{b}$. 

The quiver and all interactions can be written explicitly but we will not need the explicit matter content in the following. The reader can find many examples in \cite{Butti:2005vn,Butti:2005ps}. The only important information is that there is a very simple formula 
for the central charge of the CFT in the large $N$ limit.  According to $a$-maximization \cite{Intriligator:2003jj},  the exact central charge $a$ of the SCFT can be obtained by extremizing the trial central  charge\footnote{In general $a=\frac{9}{32} \Tr R^3 -\frac{3}{32} \Tr R^3$, but we work in the large $N$ limit where $c=a$ \cite{Henningson:1998gx}. In particular, for all our quivers, in the large $N$ limit, $\Tr R= -16 (c-a)=0$.}
\bea a(\Delta_a) =\frac{9}{32} \Tr R(\Delta_a)^3 \, ,\eea
where the trace runs over all the fermions of the theory and  $R(\Delta)$ denotes their R-charges as a function of $\Delta_a$. Explicitly, we have 
\bea\label{tHoof:linear:cubic:anomalies}
a(\Delta_a) =\frac{9}{32} N^2 \bigg(  |G| + \sum _{\Phi_{ab}} \text{ mult}(\Phi_{ab}) \left( \Delta_{\Phi_{ab}} - 1 \right)^{3} \bigg) \, ,
\eea
where $\text{mult}(\Phi_{ab})=|(e_1,v_a,v_b)|$.
 It has been shown in \cite{Benvenuti:2006xg} that the trial central charge of 
the theory can be written in the large $N$ limit as
\bea a(\Delta_a)  = \frac{9}{32}  \sum_{a,b,c=1}^d c_{abc} \Delta_a \Delta_b \Delta_c \, ,\eea
where the t'Hooft anomaly coefficients are given by 
\bea c_{abc} =\frac{N^2}{2}  |(v_a,v_b,v_c)| \, .\eea
We can remove the absolute value if we assume an order in the toric diagram. Assuming that the vertices of the toric diagram  are numerated in counter-clockwise direction, if $1\le a< b<c\le d$ we can write  $c_{abc} = N^2 (v_a,v_b,v_c)/2 >0$. The trial central charge can be then written as 
\bea
 \label{trial-a}
 a(\Delta_a) = \frac{27}{32} N^2 \sum_{1\le a<b<c\le d}  (v_a,v_b,v_c) \Delta_a \Delta_b \Delta_c \, .
\eea
 All our formulae are strictly valid in the large $N$ limit where $c=a$. 
 
 The extremization of \eqref{trial-a} gives the {\it exact} R-charges $\bar \Delta_a$ of the fields in the SCFT. They can be compared with the predictions for the dimension of baryonic operators in the gravity dual.  The baryonic operator $\det \Phi_{a-1,a}$  is obtained by wrapping a D3-branes on the three-cycle $S_a$ and the  R-charge of $\Phi_{a-1,a}$  can be computed by the standard formula \cite{Gubser:1998fp}
 \bea\label{R-chargevolume}    \bar \Delta_a = \frac{\pi \text{Vol} (S_a)}{3 \text{Vol} (Y_5)} \, .\eea
 The value of the {\it exact} central charge of the CFT is also given by \cite{Gubser:1998vd}
 \bea
  \label{avolume}
  a(\bar \Delta_a)  = \frac{\pi^3 N^2}{4 \text{Vol} (Y_5)} \, .
 \eea 
 That $a$-maximization reproduces these formulae has been tested in many examples and it can be proved in general for all toric quivers \cite{Butti:2005vn}.

\subsection{Twisted compactification to two dimensions}
\label{2dSCT}
Let us now consider   the theory compactified on a Riemann surface $\Sigma_{\fg}$ with a topological twist  and assume that it flows to a two-dimensional $\cN = (0,2)$ CFT at low energies. The gravitational dual is a type IIB solution interpolating between AdS$_5\times Y_5$ and a warped compactification AdS$_3\times_W Y_7$ where $Y_7$ is topologically a fibration of $Y_5$ over $\Sigma_\fg$ \cite{Benini:2012cz,Benini:2013cda}.    In general, we  have a family of such two-dimensional CFTs labeled by the magnetic flux of the R-symmetry on $\Sigma_{\fg}$. Once again, we can parameterize the R-symmetry flux with integers  $\fn_a$ associated with the
vertices of the toric diagram and satisfying
\bea \sum_{a=1}^d \fn_a = 2- 2\fg \, .\eea
We will refer to this constraint as the twisting condition.  It is equivalent to the requirement that the background for the R-symmetry  cancels the spin connection. As shown in  \cite{Benini:2012cz,Benini:2013cda}  the right-moving central charge of the two-dimensional theory can be found by extremizing the trial right-moving central charge
\bea
 c_r(\Delta_a,\fn_a) = 3 \Tr \gamma_3 R(\Delta_a)^2 \, ,
\eea
where $\gamma_3$ is the chirality operator in two dimensions and the trace runs over all the two-dimensional fermions. 
As shown in \cite{Hosseini:2016cyf},
the trial right-moving central charge of the theory compactified on $\Sigma_\fg$ in the large $N$ limit can be compactly written in terms of the four-dimensional trial $a$ charge as%
\footnote{Notice that in \eqref{trial-c} we impose the constraint $\sum_a \Delta_a=2$ after differentiation. For the logic behind it see appendix \ref{c versus a1}.}
\bea
 \label{trial-c}
 c_r(\Delta,\fn) & = -\frac{32}{9} \sum_{a=1}^d \fn_a \frac{\partial a(\Delta)}{\partial \Delta_a} =-3  \sum_{a,b,c=1}^d   c_{abc} \fn_a \Delta_b \Delta_c  \\
 & = 3 N^2 \sum_{1\le a<b<c\le d} (v_a,v_b,v_c)  ( \fn_a \Delta_b \Delta_c + \fn_b \Delta_a \Delta_c + \fn_c \Delta_a \Delta_b ) \, .
\eea
This relation between $c_r$ and $a$  has been proven in  \cite{Hosseini:2016cyf}   for a large class of quivers, including the toric ones,  by comparing the four-dimensional and two-dimensional central charges. It has been also verified
%even if not necessary
in many toric examples in \cite{Amariti:2017iuz}. It can be also obtained in a simple way by integrating
the four-dimensional anomaly polynomial on $\Sigma_{\fg}$, following the logic  in appendix C of \cite{Hosseini:2018uzp}. We review the derivation in appendix \ref{c versus a}. The formula is valid in the large $N$ limit where $c_r=c_l=c$. 

That $c$-extremization correctly reproduces the central charge predicted by the gravitational dual has been tested in many examples \cite{Benini:2012cz,Benini:2013cda,Benini:2015bwz,Amariti:2016mnz,Amariti:2017cyd,Amariti:2017iuz}.

\section[\texorpdfstring{$c$}{c}-extremization equals its gravity dual for all toric quivers]{$c$-extremization equals its gravity dual for all toric quivers}
\label{sec:c}

In this section we first briefly review the equivalence of $a$-maximization with the volume minimization proposed in \cite{Martelli:2005tp,Martelli:2006yb} and then we extend it
to the  equivalence of $c$-extremization with the construction proposed in  \cite{Couzens:2018wnk,Gauntlett:2018dpc} for all toric quiver. The technical parts of the proof are discussed in  appendix \ref{proof}.

\subsection[\texorpdfstring{$a$}{a}-maximization is volume minimization]{$a$-maximization is volume minimization}

The gravity dual of $a$-maximization has been found in \cite{Martelli:2005tp,Martelli:2006yb} by defining a class of off-shell backgrounds that solve the conditions for supersymmetry but relax the equations of motion. In particular, the authors of \cite{Martelli:2005tp,Martelli:2006yb} replace the Sasaki-Einsten metric on $Y_5$ with a general Sasaki metric.
The metric  depends on a  Reeb vector which  is a linear combinations of the vector fields $\partial_{\phi_i}$ generating the toric $\U(1)^3$ action
\bea
 \zeta =\sum_{i=1}^3 b_i \partial_{\phi_i} \, ,
\eea
and  specifies the direction of the R-symmetry vector field inside the isometries of $Y_5$. Supersymmetry requires $b_1=3$. 
The volumes of Sasaki manifold $Y_5$ and of its three-cycles $S_a$ are now  functions of the Reeb vector $b=(b_1,b_2,b_3)$ 
\bea
 \label{volumeSasaki}
 &\text{Vol}(Y_5) = \frac{\pi^3}{b_1} \sum_{a=1}^d \frac{ (v_{a-1},v_{a},v_{a+1})}{(v_{a-1},v_{a},b)(v_{a},v_{a+1},b)} \, , \\
 & \text{Vol}(S_a) = 2 \pi^2  \frac{ (v_{a-1},v_{a},v_{a+1})}{(v_{a-1},v_{a},b)(v_{a},v_{a+1},b)}  \, .
\eea
As shown in \cite{Martelli:2005tp,Martelli:2006yb}, the extremization of the function
\bea a(b_i) = \frac{\pi^3 N^2}{4 \text{Vol}(Y_5)} \, ,\eea
reproduces the Reeb vector $\bar b=(\bar b_1,\bar b_2,\bar b_3)$  and the volumes of the Sasaki-Einstein manifold. By construction, $a(\bar b_i)$ reproduces the gravitational prediction \eqref{avolume} for the exact central charge of the CFT.

The equivalence of $a$-maximization with  volume minimization has been proved for all toric quivers in \cite{Butti:2005vn}. The proof has been  simplified in \cite{Lee:2006ru} and generalized to other quivers in \cite{Eager:2010yu}.
Following  \cite{Butti:2005vn}, we define a {\it natural parameterization} for the R-charges in terms of the Reeb vector inspired by \eqref{R-chargevolume} 
\bea
 \label{4dparamet}
 \Delta_a(b_i) = \frac{\pi \text{Vol} (S_a)}{b_1 \text{Vol} (Y_5)} \, ,
\eea
where now the volumes are the functions of $b_i$ given  in \eqref{volumeSasaki}. Notice that $\sum_{a=1}^d \Delta_a(b_i) =2$.  One then proves that \cite{Butti:2005vn,Lee:2006ru}
\bea a(b_i)
 \label{a:toric:general}
 \equiv a (\Delta_a) \Big |_{\Delta_a(b_i)}
 \equiv \frac{27}{32} N^2 \sum_{1\le a<b<c\le d}  (v_a,v_b,v_c) \Delta_a \Delta_b \Delta_c \Big |_{\Delta_a(b_i)} \, .
\eea
  
One might be puzzled by the fact that $a$-extremization is performed on $d-1$ independent parameters, while volume minimization is an extremization with respect to the Reeb vector that depends on two independent parameters only. The Reeb vector in a sense only sees the mixing of the R-symmetry with the mesonic  symmetries. The point  is that, as proved in  \cite{Butti:2005vn}, the trial $a$-function $a (\Delta_a)$ is {\it automatically extremized}
\bea\label{id2:a}
\sum_a  B_a \frac{\partial a (\Delta_a)}{\partial \Delta_a}\Big |_{\Delta_a(b)} \equiv 0 \, ,\eea
with respect to the baryonic directions, defined by \eqref{baryonicsymm}.

\subsection[\texorpdfstring{$c$}{c}-extremization is equivalent to its gravity dual]{$c$-extremization is equivalent to its gravity dual}

The gravity dual of $c$-extremization has been found in \cite{Couzens:2018wnk,Gauntlett:2018dpc}.  The solution associated with a twisted compactification of the four-dimensional CFT on $\Sigma_\fg$ is a  warped background AdS$_3\times_W Y_7$ where $Y_7$ is topologically a fibration of $Y_5$ over $\Sigma_\fg$  with a five-form flux. The authors \cite{Couzens:2018wnk,Gauntlett:2018dpc} define a family of off-shell backgrounds, depending on the Reeb vector,  that solve the conditions for supersymmetry but relax the equations of motion for the five-form. They also define a functional of the Reeb vector that, upon extremization, selects the on-shell R-symmetry and it becomes equal to the exact two-dimensional central charge. Here we describe the basic ingredients of the construction and we refer to \cite{Couzens:2018wnk,Gauntlett:2018dpc} for details. We will work under the assumption that the gravity background associated with $Y_5$ exists. This is not always the case, as discussed in \cite{Couzens:2018wnk,Gauntlett:2018dpc}.

The off-shell backgrounds  depend on a  Reeb vector  $b=(b_1,b_2,b_3)$, and on $d$ parameters $\lambda_a$ and $d$ fluxes $\fn_a$.
The Reeb vector is again given by 
\bea \zeta =\sum_{i=1}^3 b_i \partial_{\phi_i} \, ,\eea
and  specifies the direction of the R-symmetry vector field inside the isometries of $Y_5$. This time supersymmetry requires $b_1=2$.
The parameters $\lambda_a$ are associated with the toric divisors $D_a$ and  determine the K\"ahler class of a four-dimensional transverse slice. For simplicity, we restrict to  the quasi-regular case where the quotient with respect to the Reeb action, $V=Y_5/\U(1)$, is  a four-dimensional compact toric orbifold. Then the K\"ahler class of $V$ is given by
\bea\label{kal} \omega = -2 \pi \sum_{a=1}^d \lambda_a c_a \, ,\eea
where $c_a$ are the Poincar\'e dual of the restriction of $D_a$ to $V$. Only $d-2$ parameters $\lambda_a$ are independent, since there are only $d-2$ independent two-cycles in $V$ (one more than in $Y_5$).
 We can recover the Sasaki geometry  for $\lambda_a=-1/2b_1$.\footnote{In the Sasaki case, the contact form $\eta$ associated with the Reeb vector $b$ satisfies $\rd\eta = 2\omega$ and $b_1=3$ \cite{Martelli:2005tp,Martelli:2006yb}.  Here instead    $[\rd\eta] = [\rho]/b_1$ in cohomology, where the Ricci form is given by  $\rho = 2 \pi \sum_{a=1}^d  c_a$ and $b_1=2$.}  The fluxes $\fn_a$  are also   associated with the divisors $D_a$ and satisfy the twisting condition
\be
 \sum_{a=1}^d \fn_a= 2- 2 \fg \, .
\ee
The $\fn_a$ are magnetic fluxes for both the three gauge fields associated with  the isometries of $Y_5$ and the $d-3$ gauge fields coming from the reduction of the four-form potential on the $d-3$ independent three-cycles $S_a$. The fluxes associated with  the  isometries enter explicitly  in the fibration of SE$_5$ over $\Sigma_\fg$ and they can be parameterized by the integers $n^i = \sum_{a=1}^d v_a^i \fn_a$.  They are associated with the mesonic symmetries of the quiver. The other $d-3$ fluxes enter  in the supergravity five-form and are associated with the baryonic symmetries. 
The relation with the fluxes defined in \cite{Gauntlett:2018dpc} is $M_a=-\fn_a N$.

Following  \cite{Gauntlett:2018dpc}, we define the master volume of the five-manifold  with K\"ahler class \eqref{kal} 
\bea
 \label{mastervol}
 {\cal V}=4 \pi^3 \sum_{a=1}^d \lambda_a  \frac{ \lambda_{a-1} (v_a,v_{a+1},b) - \lambda_{a} (v_{a-1},v_{a+1},b) + \lambda_{a+1} (v_{a-1},v_{a},b)}{  (v_{a-1},v_{a},b) (v_{a},v_{a+1},b)} \, .
\eea
Notice that we identify indices modulo $d$ so that $v_{d+1}=v_1$ and $\lambda_{d+1}=\lambda_1$. 
%As shown in \cite{Gauntlett:2018dpc} and reviewed in section \ref{simplifications}, $\cV$  is a function of only $d-2$ independent parameters $\lambda_a$. 
%The volumes of the Sasaki five-manifold (and its cycles $S_a$) corresponding to  $\lambda_a=-1/2b_1$  can be extracted from $\cV$ and are given by
%\bea &\text{Vol}(Y_5) = \frac{\pi^3}{b_1} \sum_{a=1}^d \frac{ (v_{a-1},v_{a},v_{a+1})}{(v_{a-1},v_{a},b)(v_{a},v_{a+1},b)} =\frac{1}{8 b_1^2} \sum_{a,b=1}^d \frac{\partial^2 {\cal V}}{\partial\lambda_a\partial \lambda_b} \, , \\
%& \text{Vol}(S_a) = 2 \pi^2  \frac{ (v_{a-1},v_{a},v_{a+1})}{(v_{a-1},v_{a},b)(v_{a},v_{a+1},b)} = \frac{1}{4 \pi  b_1}  \sum_{b=1}^d \frac{\partial^2 {\cal V}}{\partial\lambda_a\partial \lambda_b} \, .
% \eea
The supersymmetry and flux quantization conditions for the off-shell background can be then summarized by \cite{Gauntlett:2018dpc}%
\footnote{In the notation of  \cite{Gauntlett:2018dpc}, we set  $L^4=2 (2\pi l_s)^4 g_s$. In order to compare with \cite{Gauntlett:2018dpc} one must also set $\Delta_a= \frac{R_a}{N}$ and  $\fn_a = -\frac{M_a}{N}$.} 
\bea
 \label{constraint}
 & N= - \sum_{a=1}^d \frac{\partial {\cal V}}{\partial\lambda_a} \, , \\
 & \fn_a N = -\frac{A}{2\pi} \sum_{b=1}^d \frac{\partial^2 {\cal V}}{\partial\lambda_a\partial \lambda_b} - b_1 \sum_{i=1}^3 n^i  \frac{\partial^2 {\cal V}}{\partial\lambda_a\partial b_i} \, ,\\
 & A \sum_{a,b=1}^d \frac{\partial^2 {\cal V}}{\partial\lambda_a\partial \lambda_b} = 2\pi n^1 \sum_{a=1}^d \frac{\partial {\cal V}}{\partial\lambda_a} -2 \pi b_1  \sum_{i=1}^3 n^i \sum_{a=1}^d  \frac{\partial^2 {\cal V}}{\partial\lambda_a\partial b_i} \, ,
\eea
where $n^i = \sum_{a=1}^d v_a^i \fn_a$.  As shown in \cite{Gauntlett:2018dpc} and reviewed in section \ref{simplifications}, $\cV$ is a function of only $d-2$ independent parameters $\lambda_a$ and only $d-1$ equations in \eqref{constraint} are independent.  We can use the  constraints \eqref{constraint}  to eliminate  the $d-2$  independent $\lambda_a$ and $A$ and write them as functions of $b_i$ and $\fn_a$.\footnote{The dependence on the remaining two variables $\lambda_a$ drops out from every physical quantity. To simplify the computation,
one can also choose a gauge like in appendix \ref{sec:gauge}.} We then obtain the $c$-functional \cite{Gauntlett:2018dpc}
\bea
 \label{Ssusy}
 c (b_i,\fn_a) = - 48 \pi^2 \bigg ( A   \sum_{a=1}^d \frac{\partial {\cal V}}{\partial\lambda_a} +2 \pi b_1 \sum_{i=1}^3 n^i  \frac{\partial {\cal V}}{\partial b_i} \bigg ) \bigg |_{\lambda_a(b,\fn), \, A(b,\fn)}  \, .
\eea
For further reference, we  also define the on-shell value of the master volume
 \bea\label{masteronshell}  \cV_{\text{on-shell}}(b_i,\fn_a) = \cV \Big |_{\lambda_a(b,\fn), \, A(b,\fn)} \, .\eea

The authors of \cite{Gauntlett:2018dpc} checked   that the extremization of $c$ with respect to $b_i$ (with $b_1=2$) correctly reproduces the central charge of the two-dimensional CFT in various examples, including the Y$^{p,q}$ and X$^{p,q}$ manifolds. By an explicit computation along the lines of  \cite{Benini:2012cz,Benini:2013cda,Benini:2015bwz}, they also show that the identification holds off-shell and $c(b_i,\fn_a)$ can be identified with the trial right-moving central charge. We want now to show that this holds for all toric quiver, using the expression \eqref{trial-c} derived in \cite{Hosseini:2016cyf} and a {\it natural parameterization} of the R-charges based on the toric data.

In order to prove this, in analogy with \cite{Butti:2005vn} and the four-dimensional case,  we define  
\bea
 \label{BZ}
 \Delta_a(b_i,\fn_a) =  - \frac{2}{N} \frac{\partial {\cal V}}{\partial\lambda_a} \Big |_{\lambda_a(b,\fn), \, A(b,\fn)} \, ,
\eea
satisfying $\sum_a \Delta_a = 2$. This expression is indeed the holographic prediction for the R-charges of baryonic operators obtained by wrapping D3-branes on the cycles associated
with the toric divisors $D_a$ \cite{Couzens:2017nnr,Gauntlett:2018dpc}. As such this is the natural generalization of the four-dimensional parameterization \eqref{4dparamet}.
It is important to observe that the $\Delta_a$ satisfy  \cite{Gauntlett:2018dpc}
\bea
 \label{sum}
 \sum_{a=1}^d  \Delta_a v_a  = 2 \frac{b}{b_1} \Big |_{\Delta_a(b,\fn)} \, .
\eea 
We will show how to obtain an explicit expression for $\Delta_a(b_i,\fn_a)$ in section \ref{sec:examples}.

With these definitions, we will  show that\footnote{In this and the following equations we set $b_1=2$. Reinstating $b_1$ we have   $\cV =  \frac{b_1}{108 \pi^3 } a$ and $c = \frac{ b_1}{2 } c_{r}$.}
\bea
 \label{id1}
  c(b_i,\fn_a)   \equiv   c_{r} (\Delta_a, \fn_a) \Big |_{\Delta_a(b,\fn)} \equiv
 - \frac{ 32 }{9}  \sum_{a=1}^d \fn_a \frac{\partial a (\Delta_a)}{\partial \Delta_a} \Big |_{\Delta_a(b,\fn)} \, ,
\eea
thus proving the off-shell equivalence of $c$-extremization and the formalism of  \cite{Gauntlett:2018dpc}  for all toric quivers.

As  in four dimensions, one might be puzzled by the fact that $c$-extremization is performed on $d-1$ independent parameters, while the construction in  \cite{Gauntlett:2018dpc} is an extremization with respect to the Reeb vector that depends on only two independent parameters. The point  is again that 
 the trial $c$-function $c_r (\Delta_a, \fn_a)$ is {\it automatically extremized}
\bea
 \label{id2}
 \sum_a  B_a \frac{\partial c_{r} (\Delta, \fn)}{\partial \Delta_a}\Big |_{\Delta_a(b,\fn)} = \sum_{a,b}  B_a \fn_b \frac{\partial^2 a  (\Delta_a)}{\partial \Delta_a \partial \Delta_b}\Big |_{\Delta_a(b,\fn)} \equiv 0 \, ,
\eea
with respect to the baryonic directions, defined by \eqref{baryonicsymm}, as we will show. Again, this is completely analogous to \cite{Butti:2005vn}.
 
Indeed, \eqref{id1} and \eqref{id2}  follow at once from the result in appendix \ref{proof}, where  we will  prove that there exists a vector $t$ such that 
\bea \label{vector:u:constraint}  -6 \sum_{b,c=1}^d c_{abc} \fn_b \Delta_c   =  c_{r} (\Delta,\fn) + (e_1, r_a, t)  \Big |_{\Delta_a(b,\fn)}\, ,\eea
where  $r_a=  v_a - b/b_1$.   \eqref{id1} follows by multiplying \eqref{vector:u:constraint} by $\Delta_a$ and summing over $a$. The term with the vector $t$ cancels since 
$\sum_a \Delta_a r_a = 0$, as a consequence of \eqref{sum}. Similarly,  \eqref{id2} follows by multiplying \eqref{vector:u:constraint} by $B_a$ and summing over $a$. The term with $c_r$ on the right hand side vanishes because $\sum_a B_a=0$ and the term with $t$ because  $\sum_a B_a r_a =0$, where we used \eqref{baryonicsymm}. 

Let us also observe that, quite interestingly, the on-shell value of the master volume coincides with the four-dimensional trial central charge
\bea\label{id3}
 & \cV_{\text{on-shell}}(b_i,\fn_a)   \equiv  \frac{1}{54 \pi^3} a (\Delta_a) \Big |_{\Delta_a(b,\fn)} \, . \\
\eea

A word of caution is in order. To find the exact right-moving central charge of the two-dimensional CFT we need to extremize $c(b_i,\fn_a)$ with respect to $b_2$ and $b_3$ after setting $b_1=2$, or equivalently, $c(\Delta_a,\fn_a)$ with respect to $\Delta_a$ with the constraint $\sum_a \Delta_a=2$. Our results guarantee that the two procedures are equivalent for all toric quivers. However they do not guarantee that the exact central charge found in this way 
really corresponds to an IR CFT. Similarly, in gravity, nothing guarantee that the family of backgrounds  discussed in \cite{Couzens:2018wnk,Gauntlett:2018dpc} contains an actual solution of the equations of motion
of type IIB.  Explicit examples of possible obstructions are discussed in \cite{Couzens:2018wnk}.

\section{Formulae for the R-charges and examples}\label{sec:examples}
 
In this section we  discuss how to solve  equations \eqref{constraint}. Fortunately, there is no need of solving explicitly \eqref{constraint} in order to write  the   R-charges $\Delta_a(b_i,\fn_a)$. Indeed there is an explicit expression for   $\Delta_a(b_i,\fn_a)$ in terms of the toric data and the fluxes $\fn_a$. Moreover, in a convenient gauge, we can also write a general  expression for the solutions $\lambda_a$ and $A$ that allows to   write $c(b_i,\fn_a)$ and $\cV(b_i,\fn_a)$. We summarize here the result referring to  appendix \ref{proof} for  the proof.
We also discuss some explicit examples.

We first show how to find   the R-charges $\Delta_a(b_i,\fn_a)$ in terms of the toric data and the fluxes $\fn_a$. A consequence of \eqref{constraint}  is the set of equations  
\bea \label{fundamentalidentitytext}
&(v_{a-1},v_a,v_{a+1}) (v_{a+1},v_{a+2},n) \Delta_{a+1}(b_i,\fn_a)- (v_a,v_{a+1},v_{a+2})(v_{a-1},v_{a},n) \Delta_a(b_i,\fn_a) \\ 
&= - \frac{2}{b_1} \left( (v_{a-1},v_a,v_{a+1})  (v_{a+1},v_{a+2},b) \fn_{a+1} - (v_a,v_{a+1},v_{a+2})(v_{a-1} ,v_{a},b) \fn_a\right ) \, ,
\eea
where, as usual, we identify the indices modulo $d$. These equations allow to find explicitly $\Delta_a(b_i,\fn_a)$ by recursion. We can use them to 
express $\Delta_a$ in terms of $\Delta_1$, and, finally, determine $\Delta_1$ using the constraint $\sum_{a=1}^d \Delta_a =2$.

In order to write $c(b_i,\fn_a)$ and $\cV(b_i,\fn_a)$ we also need to solve \eqref{constraint} for the variables $\lambda_a$ and $A$. As already mentioned, only
$d-2$ variables $\lambda_a$ are independent. Indeed, the master volume \eqref{mastervol} is a quadratic form in $\lambda_a$ invariant under 
\bea
 \label{gaugetext }
 \lambda_a \rightarrow \lambda_a + \sum_{i=2}^3 l_i (b_1 v_a^i - b_i) \, ,
\eea
for arbitrary functions $l_2$ and $l_2$. We can use this freedom to choose a gauge, for example $\lambda_1=\lambda_2=0$. In this gauge, we can explicitly invert the relation \eqref{BZ} and write
\bea
  \label{inversiontext} \lambda_a = -\frac{N}{16\pi^3}  \sum_{c=2}^{a} (v_c,v_a,b) \Delta_c(b_i,\fn_a) \, ,\qquad a=3,\ldots ,d \, .
 \eea
 In this gauge, equations \eqref{constraint} also imply
 \bea \label{areatext} A= -\frac{N}{8\pi^2} \frac{(v_2,v_3,n) (v_1,v_2,b) }{(v_1,v_2,v_3)}  \Delta_2(b_i,\fn_a)  - \frac{N}{4\pi^2 b_1} \frac{(v_2,v_3,b) (v_1,v_2,b) }{(v_1,v_2,v_3)} \fn_2   \, .\eea  
Finally, the $c$-functional can be simplified to 
\bea
 \label{cfuncttext}
 c(b_i,\fn_a) = 48 \pi^2 N \bigg (\frac{A }{2} + \pi \sum_{a=1}^d \lambda_a \fn_a \bigg )\, .
\eea

As clear from the previous formulae and manifest in the following examples, the gauge invariant quantities $\Delta_a(b_i,\fn_a)$, $\cV(b_i,\fn_a)$ and $c(b_i,\fn_a)$  are homogeneous  polynomials  of $b_i/b_1$. In particular, $\Delta_a$ is  a linear homogeneous polynomial in $b_i/b_1$, $c_{r}/b_1$ is quadratic and $\cV/b_1$ is cubic. In the gauge  where two $\lambda_a$ are set to zero, also $\lambda_a/b_1$ and $A/b_1$ are quadratic homogeneous polynomials of $b_i/b_1$.  Setting $b_1=2$, as required by supersymmetry,  $\Delta_a$ becomes a  linear function of $b_2$ and $b_3$ with rational functions of the fluxes as coefficients. This should be contrasted with the case of $a$-maximization  \cite{Butti:2005vn}, 
where the R-charges $\Delta_a(b_i)$  are rational functions of $b$ with poles on the sides of the toric diagram, as one can see from \eqref{4dparamet} and \eqref{volumeSasaki}.

We now present few examples.  

\subsection[\texorpdfstring{$\cN = 4$}{N=4} SYM]{$\cN = 4$ SYM }

Our first example is the $\cN = 4$ super Yang-Mills (SYM) theory compactified on $\Sigma_\fg$. The holographic  dual has been found in \cite{Benini:2013cda}.
The manifold is $Y_5=S^5$ and the toric cone is specified by the vectors
\be
 \vec{v}_1 = ( 0 , 0 ) \, , \qquad \vec{v}_2 = ( 1 , 0 ) \, , \qquad \vec{v}_3 = ( 0 , 1 ) \, .
\ee
 In $\cN = 1$ notation, the four-dimensional theory contains three adjoint chiral fields $\Phi_{a}$, $a = 1,2,3$, with superpotential
\be
 \label{SYM:superpotential}
 W = \Tr \left( \Phi_3 [ \Phi_1 , \Phi_2 ] \right) \, .
\ee
In this example, the vertices are in one-to-one correspondence with the fields and fluxes. The vertex $v_a$ is associated with the field $\Phi_a$ with R-charge $\Delta_a$  and the flux $\fn_a$. They  satisfy
\be
 \label{SYM:Delta:flux:constraint}
 \sum_{a = 1}^{3} \Delta_a = 2 \, , \qquad \sum_{a = 1}^3 \fn_a = 2 - 2 \fg \, ,
\ee
which just express the fact that the superpotential \eqref{SYM:superpotential} has R-charge two. Since $d=3$, there are no baryonic symmetries. 
The trial  central charge $a$, at large $N$, reads (\cf\;\eqref{tHoof:linear:cubic:anomalies} and \eqref{trial-a})
\be
 a (\Delta_a) = \frac{9}{32} N^2 \bigg( 1 + \sum_{a=1}^3 (\Delta_a-1)^3 \bigg) = \frac{27}{32} N^2 \Delta_1 \Delta_2 \Delta_3 \, .
\ee
The trial  central charge $c_r$ is given by \eqref{trial-c} 
\bea 
 c_r ( \Delta_a , \fn_a ) = -3 N^2 ( \Delta_1 \Delta_2 \fn_3 + \Delta_2 \Delta_3 \fn_1 + \Delta_1 \Delta_3 \fn_2 )
\, .
\eea 

Solving explicitly \eqref{constraint} or using the recursion relations \eqref{fundamentalidentitytext} we find 
\be
 \label{SYM:Delta:b}
 \Delta_1(b_i) = \frac{2}{b_1} ( b_1 - b_2 - b_3 ) \, , \qquad \Delta_2(b_i) = \frac{2 b_2}{b_1} \, , \qquad \Delta_3(b_i) = \frac{2 b_3}{b_1} \, .
\ee
Notice that $\Delta_a(b_i)$ are independent of the fluxes $\fn_a$. This is due to the absence of baryonic symmetries. Moreover, comparing with \eqref{volumeSasaki} 
we see that 
\bea
 \Delta_a (b_i) \equiv  \frac{\pi}{b_1} \frac{\text{Vol}(S_a)}{\text{Vol}(Y_5)} \, .
\eea
Therefore, for ${\cal N}=4$ SYM the parameterization \eqref{BZ} coincides with the one used in \cite{Butti:2005vn} for $a$-maximization. Moreover, the on-shell value of the master volume
is given by
\bea\label{masterN4} \cV_{\text{on-shell}} (b_i)= \frac{N^2}{16 b_1^2 \text{Vol}(S^5)} \, ,\eea
and again is independent of $\fn_a$. Setting $b_1=2$ we find the very simple identification
\bea \Delta_1 =2-b_2-b_3\, ,\qquad \Delta_2=b_2\, ,\qquad  \Delta_3=b_3 \, .\eea

One can easily verify that \eqref{id1} and \eqref{id3} are satisfied. For ${\cal N}=4$ SYM, \eqref{id3} is just equivalent to the equivalence of $a$-maximization and volume minimization found in  \cite{Butti:2005vn},
since \eqref{masterN4} holds and the parameterization of R-charges is the same in two and four dimensions.  

\subsection{Klebanov-Witten theory}

Our second example is the twisted compactification of the Klebanov-Witten theory \cite{Klebanov:1998hh} on $\Sigma_\fg$, discussed \eg\;in \cite{Benini:2015bwz,Hosseini:2016cyf}.
The manifold in this case is $Y_5 = T^{1,1}$.
%that can be described as the coset $\SU(2) \times \SU(2) / \U(1)$.
The toric cone $C(T^{1,1})$ is determined by the vectors
\be
 \vec{v}_1 = ( 0 , 0 ) \, , \qquad \vec{v}_2 = ( 1 , 0 ) \, , \qquad \vec{v}_3 = ( 1 , 1 ) \, , \qquad \vec{v}_4 = ( 0 , 1 ) \, .
\ee
This theory has $\cN=1$ supersymmetry. The quiver contains two $\SU(N)$ gauge groups with two bi-fundamental chiral fields $A_i$ in the representation $({\bf N},\overline{{\bf N}})$ and two bi-fundamental chiral fields $B_i$ in the representation $(\overline{{\bf N}},{\bf N})$.
The theory has a quartic superpotential
\be
 \label{ABJMsuper}
 W = \Tr \big( A_1B_1A_2B_2 - A_1B_2A_2B_1 \big) \, .
\ee
We introduce four chemical potentials $\Delta_a$ and fluxes $\fn_a$, one for each of the four fields $\{A_i,B_i\}$, associated with the four
 vertices $v_a$, with the constraints
\be
 \label{KW:Delta:flux:constraint}
 \sum_{a = 1}^{4} \Delta_a = 2 \, , \qquad \sum_{a = 1}^4 \fn_a = 2 - 2 \fg \, .
\ee
The trial $a$ central charge can be computed from either \eqref{tHoof:linear:cubic:anomalies} or \eqref{trial-a} and it reads
\be
 a (\Delta_a) = \frac{27}{32} N^2 (\Delta_1 \Delta_2 \Delta_3+\Delta_1 \Delta_2 \Delta_4+\Delta_1 \Delta_3 \Delta_4 +\Delta_2 \Delta_3 \Delta_4) \, .
\ee
The trial central charge $c_r$ is given by \eqref{trial-c}
\be
 c_r ( \Delta_a , \fn_a ) = - 3 N^2 \sum_{\substack{ a < b \\ ( a , b ) \neq c}}^{4} \Delta_a \Delta_b \fn_c \, .
\ee
Solving explicitly \eqref{constraint} or using the recursion relations \eqref{fundamentalidentitytext} we find
\bea
\label{RchargesKW}
 \Delta_1(b_i,\fn_a) = & \frac{2}{b_1} \frac{b_1 (\fn_1+\fn_2+\fn_4) - b_2 (\fn_1+\fn_2) - b_3 (\fn_1+\fn_4)}{\fn_1+\fn_2+\fn_3+\fn_4} \, , \\
 \Delta_2(b_i,\fn_a) = & \frac{2}{b_1} \frac{b_2 (\fn_1+\fn_2)+b_1 \fn_3-b_3 (\fn_2+\fn_3)}{\fn_1+\fn_2+\fn_3+\fn_4} \, , \\
 \Delta_3(b_i,\fn_a) = & \frac{2}{b_1} \frac{-b_1 \fn_3+b_3 (\fn_2+\fn_3)+b_2 (\fn_3+\fn_4)}{\fn_1+\fn_2+\fn_3+\fn_4} \, , \\
 \Delta_4(b_i,\fn_a) = & \frac{2}{b_1} \frac{b_1 \fn_3+b_3 (\fn_1+\fn_4)-b_2 (\fn_3+\fn_4)}{\fn_1+\fn_2+\fn_3+\fn_4} \, .
\eea
Notice that these are linear polynomials in $b_i/b_1$.

The baryonic symmetry $\U(1)_B$ is characterized by \eqref{baryonicsymm} and it is given by
\be
 B_1 =- 1 \, , \qquad B_2 = 1 \, , \qquad B_3 = - 1 \, , \qquad B_4 = 1 \, .
\ee
Thus, the decoupling condition \eqref{id2} can be explicitly written as
\be
 \label{baryonic:KW}
 \Delta_1 \fn_3 + \Delta_3 \fn_1 - \Delta_2 \fn_4 - \Delta_4 \fn_2 = 0 \, .
\ee
One can check that \eqref{baryonic:KW} is automatically satisfied by the solution \eqref{RchargesKW}. 

Finally, $c(b_i,\fn_a)$ and $\cV_{\text{on-shell}} (b_i,\fn_a)$ read
\bea
  \cV_{\text{on-shell}} (b_i,\fn_a) & = \frac{N^2}{16 \pi^3 b_1^2 (\fn_1+\fn_2+\fn_3+\fn_4)^2}
  \big[ 2 b_1^2 \fn_3 (b_3 (\fn_2+\fn_3)+b_2 (\fn_3+\fn_4)) \\
  & - b_1 \left( b_3^2 (\fn_2+\fn_3)^2 + b_2^2 (\fn_3+\fn_4)^2 + b_1^2 \fn_3^2 \right) \\
  & + b_1 b_2 b_3 \left(\fn_1^2+2 (\fn_2+\fn_4) \fn_1+\fn_2^2-3 \fn_3^2+\fn_4^2-2 \fn_2 \fn_3-2 \fn_3 \fn_4\right) \\
  & - b_2 b_3 (\fn_1+\fn_2+\fn_3+\fn_4) (b_2 (\fn_1+\fn_2-\fn_3-\fn_4)+b_3 (\fn_1-\fn_2-\fn_3+\fn_4)) \big] , \\
  c(b_i,\fn_a) & = \frac{6 N^2}{b_1 (\fn_1+\fn_2+\fn_3+\fn_4)} \big[ b_2^2 (\fn_1+\fn_2) (\fn_3+\fn_4) \\
  & - b_2 \left(b_1 (\fn_1+\fn_2-\fn_3+\fn_4) (\fn_3+\fn_4)+b_3 \left(\fn_1^2-\fn_2^2+\fn_3^2-\fn_4^2\right)\right) \\
  & - b_1^2 \fn_3^2-b_3^2 (\fn_2+\fn_3) (\fn_1+\fn_4)+b_1 b_3 (\fn_2+\fn_3) (\fn_1+\fn_2-\fn_3+\fn_4) \big] \, .
\eea
One can explicitly verify that \eqref{id1} and \eqref{id3} are satisfied. Recall that the central charge is obtained by extremizing $c(b_i,\fn_a)$ with respect to $b_2$ and $b_3$ after setting $b_2=2$.

\subsection[\texorpdfstring{$Y^{p,q}$ quiver gauge theory}{Y[p,q]}]{$Y^{p,q}$ quiver gauge theory }

Our third example is the $Y^{p,q}$ ($p>0$ and $p \geq q \geq 0$) quiver gauge theory \cite{Benvenuti:2004dy}. The dual of the twisted compactification on $\Sigma_{\fg}$ is discussed  in \cite{Benini:2015bwz}.
The cone $C(Y^{p,q})$ determines a polytope with four vertices \cite{Martelli:2004wu}
\be
 \vec{v}_1 = ( 0 , 0 ) \, , \qquad \vec{v}_2 = ( 1 , 0 ) \, , \qquad \vec{v}_3 = ( 0 , p ) \, , \qquad \vec{v}_4 = ( - 1 , p + q ) \, .
\ee
The $Y^{p,q}$ quiver has $2p$ $\SU(N)$ gauge groups with $4 p + 2 q$ chiral fields $\{Y,Z,U^\alpha,V^\alpha\}$, $\alpha = 1, 2$, in bi-fundamental representations of pairs of gauge groups. In \eqref{Yp,q:table} we present the R-charges and their multiplicity.
\begin{equation}
\label{Yp,q:table}
\begin{array}{c|c|c|c}
(a,b) \text{ in} \,  \Phi_{ab}   & \text{multiplicity} & \U(1)_R & \text{fields} \\
 \hline
 (4,1) & p + q & \Delta_1 & Y \\
 (1,2) & p & \Delta_2 & U^1 \\
 (2,3) & p - q & \Delta_3 & Z \\
 (3,4) & p & \Delta_4 & U^2 \\
 (1,3) & q & \Delta_2 + \Delta_3 & V^1 \\
 (2,4) & q & \Delta_3 + \Delta_4 & V^2 \\
\end{array}
\end{equation}
Supersymmetry imposes the constraints
\be
 \sum_{a = 1}^{4} \Delta_a = 2 \, , \qquad \sum_{a = 1}^{4} \fn_a = 2 - 2 \fg \, .
\ee
The $a$ central charge can be computed from either \eqref{tHoof:linear:cubic:anomalies} or \eqref{trial-a} and it is given by
\be
 \frac{32}{27 N^2} a (\Delta_a) = \Delta_2 ( \Delta_1 - \Delta_3 ) \Delta_4 q + \left[ \Delta_2 \Delta_3 \Delta_4 + \Delta_1 \Delta_3 \Delta_4 + \Delta_1 \Delta_2 \Delta_3+ \Delta_1 \Delta_2\Delta_4) \right] p \, .
\ee
Solving explicitly \eqref{constraint} or using the recursion relations \eqref{fundamentalidentitytext} we find 
\bea
\label{RchargesYpq}
 \Delta_1(b_i,\fn_a) = & \frac{2}{b_1} \frac{p^3 (b_1 (\fn_1+\fn_2+\fn_4)-b_2 (\fn_1+\fn_2)) + p q^2 (b_2 \fn_2 - b_1 (\fn_2+\fn_4))}{p \left((\fn_1+\fn_2+\fn_3+\fn_4) p^2+(\fn_1-\fn_3) p q-(\fn_2+\fn_4) q^2\right)} \\
 - & \frac{2}{b_1} \frac{p^2 (b_3 (2 \fn_1+\fn_2+\fn_4) - b_1 \fn_1 q + b_2 (\fn_1+\fn_4) q) - q^2 (b_3 (\fn_2+\fn_4)+b_2 \fn_4 q)}{p \left((\fn_1+\fn_2+\fn_3+\fn_4) p^2+(\fn_1-\fn_3) p q-(\fn_2+\fn_4) q^2\right)} \, , \\
 \Delta_2(b_i,\fn_a) = & \frac{2}{b_1} \frac{p^2 (b_1 \fn_3 + b_2 (\fn_1+\fn_2)) + (\fn_1-\fn_3) p (b_2 q+b_3) - b_2 \fn_2 q^2}{(\fn_1+\fn_2+\fn_3+\fn_4) p^2 + (\fn_1-\fn_3) p q - (\fn_2+\fn_4) q^2} \, , \\
 \Delta_3(b_i,\fn_a) = & \frac{2}{b_1} \frac{p^3 (b_2 (\fn_3+\fn_4)-b_1 \fn_3)+p^2 (b_3 (\fn_2+2 \fn_3+\fn_4) - b_1 \fn_3 q + b_2 (\fn_3+\fn_4) q)}{p \left((\fn_1+\fn_2+\fn_3+\fn_4) p^2+(\fn_1-\fn_3) p q-(\fn_2+\fn_4) q^2\right)} \\
 - & \frac{2}{b_1} \frac{b_2 \fn_4 p q^2 + q^2 (b_3 (\fn_2+\fn_4) + b_2 \fn_4 q)}{p \left((\fn_1+\fn_2+\fn_3+\fn_4) p^2+(\fn_1-\fn_3) p q-(\fn_2+\fn_4) q^2\right)} \, , \\
 \Delta_4(b_i,\fn_a) = & \frac{2}{b_1} \frac{p^2 (b_1 \fn_3-b_2 (\fn_3+\fn_4))+b_3 (\fn_1-\fn_3) p+b_2 \fn_4 q^2}{(\fn_1+\fn_2+\fn_3+\fn_4) p^2+(\fn_1-\fn_3) p q-(\fn_2+\fn_4) q^2} \, .
\eea 
Notice that these are linear polynomials in $b_i/b_1$. At the end of the computation we can set $b_1=2$ as required by supersymmetry.

The baryonic symmetry $\U(1)_B$ is characterized by \eqref{baryonicsymm}. It reads
\be
 B_1 =- \frac{p-q}{p} \, , \qquad B_2 = 1 \, , \qquad B_3 = - \frac{p+q}{p} \, , \qquad B_4 = 1 \, .
\ee
Hence, the decoupling condition \eqref{id2} can be explicitly written as
\be
 \Delta_1 \fn_2 p + \Delta_3 \fn_1 p + \fn_2 \Delta_4 \left( \frac{q^2}{p} - p \right) + \fn_4 \Delta_2 \left( \frac{q^2}{p} - p \right) = 0 \, ,
\ee
and one can see that it is automatically satisfied by the solution \eqref{RchargesYpq}. 

The expressions for $c(b_i,\fn_a)$ and $\cV_{\text{on-shell}} (b_i,\fn_a)$ are too long to be reported here. One can explicitly verify that \eqref{id1} and \eqref{id3} are satisfied.

\section[\texorpdfstring{$\cI$}{I}-extremization and black hole entropy]{$\cI$-extremization and black hole entropy}\label{sec:BH}

As discussed in \cite{Couzens:2018wnk}, the construction behind the gravity dual of $c$-extremization can be extended to twisted compactifications of three-dimensional CFTs on $\Sigma_\fg$.
The gravity dual of the IR physics is a warped background AdS$_2 \times_W Y_9$ where $Y_9$ is topologically a fibration of a Sasaki-Einstein $Y_7$ over $\Sigma_\fg$.
The gravity dual then describes the horizon of magnetically charged black holes in AdS$_4\times Y_7$. As shown in \cite{Couzens:2018wnk} for the case of solutions
of minimal gauged supergravity in four dimensions, the extremization of the analogue of the $c$-functional \eqref{Ssusy} reproduces the entropy of the black hole. It is then natural to conjecture
that the construction in \cite{Couzens:2018wnk} is the dual of $\cI$-extremization \cite{Benini:2015eyy,Benini:2016rke}, that it has been successfully used to perform a microscopic counting for AdS$_4$ black holes.
The $\cI$-extremization principle states that the entropy of magnetically charged static black holes can be obtained by extremizing the logarithm of the supersymmetric partition function on $\Sigma_\fg\times S^1$
--- also known as topologically twisted index. The index is a function of chemical potentials and magnetic fluxes for the global symmetries of the theory \cite{Benini:2015noa,Benini:2016hjo,Closset:2016arn}.
In the case of the ABJM theory \cite{Aharony:2008ug}, where $Y_7=S^7$, the $\cI$-extremization principle states that the entropy of 
magnetically charged static black holes in AdS$_4\times S^7$ is the extremum of the function \cite{Benini:2015eyy,Benini:2016rke}
\bea
 \label{iextr}
 \cI(\Delta_a,\fn_a) = -\frac{1}{2}  \sum_{a=1}^4 \fn_a \frac{\partial F_{S^3}(\Delta_a)}{\partial \Delta_a} \, ,
\eea
where
\bea
 \label{FS3}
 F_{S^3} (\Delta_a) = \frac{4 \pi \sqrt{2}N^{3/2}}{3}\sqrt{\Delta_1\Delta_2\Delta_3\Delta_4} \, ,
\eea
and the chemical potentials and the fluxes satisfy the constraints 
\be
 \label{ABJMconstraint}
 \sum_{a = 1}^{4} \Delta_a = 2 \, , \qquad \sum_{a = 1}^4 \fn_a = 2 - 2 \fg \, .
\ee
As noticed in \cite{Hosseini:2016tor},  the function \eqref{FS3} is the free energy on $S^3$ of ABJM.

In this note we show that the construction of \cite{Gauntlett:2018dpc} adapted to three-dimensional $\cN = 2$ theories, exactly reproduces \eqref{iextr} for ABJM and the identification is valid {\it off-shell}, when we use a natural parameterization of the R-charges in terms of the Reeb vector.
The parallel with four dimensions is complete. As shown in \cite{Herzog:2010hf,Jafferis:2011zi}, the extremization of $F_{S^3}$ is equivalent to volume minimization for Calabi-Yau eight-folds \cite{Martelli:2005tp,Martelli:2006yb}.
The equivalence of the  construction of \cite{Gauntlett:2018dpc} with \eqref{iextr} is the analogue of \eqref{id1}.

It was shown in \cite{Hosseini:2016tor} that \eqref{iextr}, where $\cI$ is the logarithm of the topologically twisted index and $F_{S^3} (\Delta_a)$ the $S^3$ free energy, can be extended to many quivers dual to AdS$_4\times Y_7$.
\eqref{iextr} has been called the index theorem in \cite{Hosseini:2016tor}. This may suggest that the equivalence between the construction in \cite{Gauntlett:2018dpc} and $\cI$-extremization extends to other Sasaki-Einstein manifolds.
Since the equations are more complicated to solve for Calabi-Yau eight-folds we leave this very interesting investigation to future work. This might shed light on some puzzles about baryonic symmetries raised in \cite{Hosseini:2016tor,Hosseini:2016ume,Azzurli:2017kxo}. 

\subsection{The ABJM theory}
\label{sec:ABJM}

Let us consider the twisted compactification of ABJM on $\Sigma_{\fg}$. The ABJM theory \cite{Aharony:2008ug} is a three-dimensional supersymmetric Chern-Simons-matter theory with gauge group $\U(N)_k \times \U(N)_{-k}$
(the subscripts denote the CS levels) with two bi-fundamental chiral fields $A_i$ in the representation $({\bf N},\overline{{\bf N}})$ and two bi-fundamental chiral fields $B_i$ in the representation $(\overline{{\bf N}},{\bf N})$.
The theory has a quartic superpotential
\be
 \label{ABJMsuper}
 W = \Tr \big( A_1B_1A_2B_2 - A_1B_2A_2B_1 \big) \, .
\ee
We introduce four chemical potentials $\Delta_a$, one for each of the four fields $\{A_i,B_i\}$, and four fluxes $\fn_a$ on $\Sigma_{\fg}$ satisfying
\be
 \label{ABJMconstraint2}
 \sum_{a = 1}^{4} \Delta_a = 2 \, , \qquad \sum_{a = 1}^4 \fn_a = 2 - 2 \fg \, .
\ee
The ABJM theory in three dimensions is dual to AdS$_4\times S^7/\mathbb{Z}_k$. We are interested in $k=1$, which corresponds to the toric Calabi-Yau four-fold $\mathbb{C}^4$.
The toric data are
\bea
 v_1 = (1,0,0,0) \, , \qquad v_2 = (1,1,0,0) \, , \qquad v_3 = (1,0,1,0) \, , \qquad v_4 = (1,0,0,1) \, .
\eea
We can associate each vertex to one of the fields. The supergravity background corresponding to the twisted compactification is then a warped background AdS$_2\times_W Y_9$ where $Y_9$ is topologically a fibration of  $S^7$ over $\Sigma_\fg$. It corresponds to the horizon geometry of the black holes found and studied in \cite{Cacciatori:2009iz,DallAgata:2010ejj,Hristov:2010ri}.

The master volume in \cite{Gauntlett:2018dpc} is defined as the volume of a dual polytope associated with the K\"ahler parameter $\lambda_a$:
\bea
 \cV = \frac{(2\pi)^4}{|b|} \text{Vol} \left( \left \{ y\in H(b) \, |\, (y-y_0,v_a) \ge \lambda_a \, , a=1,\ldots, 4 \right \} \right) ,
\eea
where $H(b)$ is the hyperplane $(y,b)=1/2$ and $y_0=(1,0,0,0)/(2b_1)$. Supersymmetry now requires $b_1=1$ \cite{Couzens:2018wnk}. For $\mathbb{C}^4$ the  dual polytope is  a tetrahedron lying on $H(b)$. Its four vertices can be found by solving for every distinct triple $v_a,v_b,v_c$ the equations 
\be
 (y-y_0,v_a)= \lambda_a \, , \quad (y-y_0,v_b)= \lambda_b \, , \quad (y-y_0,v_c)= \lambda_c \, , \quad (y-y_0,b)= 0 \, ,
\ee
and the volume can be easily computed. We then find that
\be\label{masterABJM}
 \cV = \frac{8 \pi^4 ( \lambda_1 (b_2+b_3+b_4-b_1) -\lambda_2 b_2 -\lambda_3 b_3 -\lambda_4 b_4)^3}{3 b_2 b_3 b_4 (b_1-b_2-b_3-b_4)} \, .
\ee
By adapting the arguments in \cite{Gauntlett:2018dpc}, it is easy to see that equations \eqref{constraint}
have the same form with the index $i$ running from $1$ to $4$.
Following \cite{Couzens:2018wnk,Gauntlett:2018dpc} we also define the entropy functional\footnote{In the notation of  \cite{Couzens:2018wnk}, we set  $L^6=(2\pi l_P)^6$.}  
\bea
 \label{Ssusy2}
 S (b_i,\fn_a) = - 8 \pi^2 \bigg ( A   \sum_{a=1}^d \frac{\partial {\cal V}}{\partial\lambda_a} +2 \pi b_1 \sum_{i=1}^4 n^i  \frac{\partial {\cal V}}{\partial b_i} \bigg ) \bigg |_{\lambda_a(b,\fn), \, A(b,\fn)}  \, ,
\eea

The equations \eqref{constraint} are easily solved for the independent $\lambda_a$ and $A$.   By substituting the result into \eqref{masterABJM} we find that
\bea
 \cV_{\text{on-shell}} (b_i,\fn_a) = \frac{N^{3/2}}{6\sqrt{2} b_1 \pi^2} \sqrt{\frac{b_2 b_3 b_4 (b_1-b_2-b_3-b_4)}{b_1}} \, .
\eea
For the entropy functional we obtain
\bea  
 S (b_i,\fn_a) = - \frac{2 \pi \sqrt{2}N^{3/2}}{3} \sqrt{\frac{b_2 b_3 b_4 (b_1-b_2-b_3-b_4)}{b_1}}
 \left ( \frac{\fn_1}{b_1-b_2-b_3-b_4}+\frac{\fn_2}{b_2} + \frac{\fn_3}{b_3} + \frac{\fn_4}{b_4} \right ) .
\eea

Similarly to what we did for $c$-extremization, we use the following parameterization for the R-charges 
\bea
 \label{BZ2}
 \Delta_a(b_i,\fn_a) =  - \frac{2}{N} \frac{\partial {\cal V}}{\partial\lambda_a} \Big |_{\lambda_a(b,\fn), \, A(b,\fn)} \, .
\eea
Plugging the solution to \eqref{constraint} into \eqref{BZ2}
we obtain
\bea 
 \Delta_1(b_i)= \frac{2(b_1-b_2-b_3-b_4)}{b_1} \, , ~~~ \Delta_2(b_i)= \frac{2 b_2}{b_1} \, , ~~~ \Delta_3(b_i)= \frac{2 b_3}{b_1} \, , ~~~ \Delta_4(b_i)= \frac{2 b_4}{b_1} \, .
\eea
As in  ${\cal N}=4$ SYM in one dimension more (see \eqref{SYM:Delta:b}), the R-charges are functions of $b_i$ only. Moreover, one can also check that they are expressed in terms of  the Sasaki volumes
\bea
 \Delta_a (b_i) \equiv \frac{2\pi}{3b_1} \frac{\text{Vol}(S_a)}{\text{Vol}(S^7)} \, ,
\eea
of toric divisors (see \cite{Martelli:2005tp,Martelli:2006yb,Hanany:2008fj,Jafferis:2011zi} for explicit expressions).

We now easily see that%
\footnote{We  set $b_1=1$ in the following formulae. Reinstating $b_1$ we have $S= \sqrt{b_1} \cI$ and $\cV=\frac{\sqrt{b_1}}{64 \pi^3} F_{S^3}$.}
\bea\label{SABJM}
 S (b_i,\fn_a)  =   \cI(\Delta_a,\fn_a)  \Big |_{\Delta_a(b)} &=  -\frac{2 \pi \sqrt{2}N^{3/2}}{3} \sum_{a=1}^4 \fn_a \frac{\partial \sqrt{\Delta_1\Delta_2\Delta_3\Delta_4}}{\partial \Delta_a}  \Big |_{\Delta_a(b)}  \\
 &= -\frac{1}{2}  \sum_{a=1}^4 \fn_a \frac{\partial F_{S^3}(\Delta_a)}{\partial \Delta_a}  \Big |_{\Delta_a(b)}  \, ,
\eea
thus proving that the functional $S (b_i,\fn_a)$ is equivalent to the $\cI$-functional defined in \cite{Benini:2015eyy,Benini:2016rke}. It is also interesting to observe that 
\bea\label{VABJM}
\cV_{\text{on-shell}} (b_i,\fn_a) =  \frac{1}{64 \pi^3} F_{S^3} (\Delta_a) \Big |_{\Delta_a = \Delta_a(b_i)} \, .
\eea
\eqref{SABJM} and \eqref{VABJM} are the direct analogues of \eqref{id1} and \eqref{id3}.

The entropy of the black holes in \cite{Cacciatori:2009iz,DallAgata:2010ejj,Hristov:2010ri} can be found equivalently by extremizing $S (b_i,\fn_a)$ with respect to $b_2,b_3,b_4$
after setting $b_1=1$ or by extremizing  $\cI(\Delta_a,\fn_a)$ with respect to $\Delta_a$ with the constraint $\sum_a \Delta_a =2$. More interestingly,
the equivalence between the construction of \cite{Gauntlett:2018dpc} and the $\cI$-extremization principle  holds also  {\it off-shell}.  It would be interesting to see if this result extends to more general Sasaki-Einstein manifolds $Y_7$.

\section*{Acknowledgements}

We would like to thank Noppadol Mekareeya for useful discussions. The work of SMH was supported by World Premier International Research Center Initiative (WPI Initiative), MEXT, Japan.
AZ is partially supported by the INFN and ERC-STG grant 637844-HBQFTNCER.

\begin{appendix}

\section[The relation between \texorpdfstring{$c_r(\Delta_a , \fn_a)$}{c[r](Delta(a),n(a))} and \texorpdfstring{$a(\Delta_a)$}{a(Delta(a))}]{The relation between $c_r(\Delta_a , \fn_a)$ and $a(\Delta_a)$}
\label{c versus a}

In this appendix we briefly review the derivation of \eqref{trial-c} given in \cite{Hosseini:2016cyf} and give an alternative one using the integration of the anomaly polynomial,
in the spirit of appendix C of \cite{Hosseini:2018uzp}.

\subsection{Direct evaluation}\label{c versus a1}
The trial $a$ central charge of a four-dimensional $\cN = 1$ field theory with gauge group $G$, at large $N$, reads
\bea\label{tHoof:linear:cubic:anomalies2}
a(\Delta_I) = \frac{9}{32} \Tr R^{3} (\Delta_I) & = \frac{9}{32}\bigg( \text{dim }G  + \sum _{I} \text{dim }\fR_I \left( \Delta_I - 1 \right)^{3} \bigg ) \, ,
\eea
where the trace is taken over all the bi-fundamental fermions and gauginos
and $\text{dim }\fR_I$ is the dimension of the respective matter representation with R-charge $\Delta_I$.
On the other hand, the trial right-moving central charge of the IR two-dimensional $\cN = (0, 2)$ SCFT can be computed from the spectrum of massless fermions \cite{Benini:2012cz,Benini:2013cda,Benini:2015bwz}.
These are gauginos 
for all the gauge groups and  fermionic zero modes for each chiral field.
The difference between the number of fermions of opposite chiralities is predicted by the Riemann-Roch theorem and equals $\fg-1$ for gauginos and $-\fn_I-\fg+ 1$ for chiral fields \cite{Benini:2012cz,Benini:2013cda,Benini:2015bwz}.
The trial right-moving central  charge is then given by
\bea
 \label{c2d:anomaly0}
 c_{r} \left( \Delta_I , \fn_I \right)  = 3 \bigg( (\fg-1)  \text{dim }G+ \sum_{I} \text{dim }\fR_I \left(1  -\fn_I - \fg  \right) \left( \Delta_I - 1 \right)^2 \bigg)  \, .
\eea
Using \eqref{tHoof:linear:cubic:anomalies2}, it is easy to see that we can write 
\bea
\label{QFTrelation}
 c_{r} \left( \Delta_I, \fn_I \right) & = -  \frac{32}{9} (1-\fg)  \bigg (  3 a  ( \Delta_I ) + \sum_{I}  \left( \frac{\fn_I}{1-\fg} - \Delta_I  \right) \frac{\partial a ( \Delta_I )}{\partial \Delta_I} \bigg )
\, .
\eea
When   $a(\Delta_I)$ is a homogeneous function of degree three of the variables $\Delta_I$,    \eqref{QFTrelation} simplifies to
\bea
\label{QFTrelation2}
 c_{r} \left( \Delta_I, \fn_I \right) & = -  \frac{32}{9}\sum_{I} \fn_I \frac{\partial a \left( \Delta_I \right)}{\partial \Delta_I} 
\, ,
\eea   
 which is precisely \eqref{trial-c}.  

In evaluating the right hand side of  \eqref{QFTrelation}, we have considered the R-charges $\Delta_I$ of all the chiral fields as independent variables. However, the R-charges satisfy the constraint $\sum_{I \in W} \Delta_I  = 2$
for each term $W$ in the superpotential. Similarly $\sum_{I \in W} \fn_I  = 2-2\fg$. Fortunately, the differential operator in \eqref{QFTrelation} is such that we can impose the constraints equivalently before or after differentiation.
\eqref{QFTrelation} is indeed valid for all parameterizations of the R-charges and fluxes (even redundant ones) provided that, if we impose a constraint coming from a superpotential term $W$, $\sum_{I \in W} \Delta_I  = 2$, a similar constraint is imposed on  $\fn_I$, $\sum_{I \in  W} \fn_I  = 2-2\fg$. In particular, it is valid for the parameterization used in this note, where we express $d-1$ independent R-charges in terms of $d$ parameters $\Delta_a$ with a constraint $\sum_{a=1}^d \Delta_a = 2$. We can apply  \eqref{QFTrelation} and \eqref{QFTrelation2}  considering $a$ as a function of $d$ independent variables $\Delta_a$ and impose the constraint $\sum_{a=1}^d \Delta_a = 2$ after differentiation.

\subsection{Integrating the anomaly polynomial}\label{c versus a2}

The trial 't Hooft anomaly coefficients of  two-dimensional $\cN = (0,2)$ CFT can be extracted by integrating the six-form anomaly polynomial $I_6$ of
the four-dimensional $\cN=1$ field theory 
over $\Sigma_{\fg}$ \cite{Alday:2009qq,Bah:2011vv,Bah:2012dg,Benini:2013cda}.
The six-form anomaly  polynomial reads 
\be
 \label{6form}
 I_6 = \frac{\Tr (F^3)}{6}  -  \frac{p_1(T\cM)}{24}\, \Tr (F)  \, ,
\ee
where $p_1(T\cM)$ is the first Poyntryagin class of tangent bundle, $F$ is the curvature of the R- and global symmetry bundle $K$ and the trace runs over all the fermions in the theory. 
We choose  a basis of generators $T^a$ adapted to the parameterization discussed in section \ref{sec:4dSCFT} and write $F = \sum_a  \Delta_a T^a c_1(F)$,
where $c_1(F)$ is a flux coupled to the $\U(1)$ R-symmetry and $\sum_a \Delta_a=2$. We can extract the trial $a$ central charge, at large $N$, from
\be
 \label{I6:a}
 I_6 = \frac{16}{27} a(\Delta_a) c_1(F)^3 \, ,
\ee
and we  find
\bea a(\Delta_a) = \frac{9}{32} \sum_{abc} c_{abc} \Delta_a \Delta_b \Delta_c \, , \eea
where $c_{abc}=\Tr ( T^a T^b T^c)$ are the t'Hooft anomaly coefficients.  

Consider now the compactification of the four-dimensional theories on a Riemann surface $\Sigma_{\fg}$ with fluxes $\fn_a$.
The prescription in \cite{Alday:2009qq,Bah:2011vv,Bah:2012dg,Benini:2013cda} for computing the anomaly coefficient $c_r$ of the two-dimensional SCFT amounts to first replace $F$ in \eqref{6form} with
\be
 \label{replacement1}
 F \to \sum_a  \left ( \Delta_a T^a  c_1(F) -  \frac{\fn_a T^a}{2 - 2 \fg} x \right )\, ,
\ee
implementing the topological twist along $\Sigma_{\fg}$, and then integrate the $I_6$ on $\Sigma_{\fg}$:
\be
 I_4 = \int_{\Sigma_{\fg}} I_6 \, .
\ee
Here $x$ denotes the Chern root of the tangent bundle to $\Sigma_{\fg}$, $\Delta_a$ parameterize the trial R-symmetry, and $\fn_a$ are the fluxes parameterizing the twist, satisfying
\be
\sum_a  \fn_a = 2 - 2 \fg \, .
\ee
Then we integrate $I_6$ over $\Sigma_{\fg}$ using $\int_{\Sigma_{\fg}} x =2 - 2 \fg$.
The result should be compared with the four-form anomaly polynomial of the two-dimensional SCFT that, in the large $N$ limit, where $c_l=c_r$, reads
\be
 I_4 = \frac{c_r(\Delta_a ,  \fn_a)}{6}  c_1(F)^2 \, .
\ee 
We see immediately that, since in our basis $a(\Delta_a)$ is homogeneous,  
\bea
 c_r(\Delta_a , \fn_a) = -  \frac{32}{9}\sum_{a} \fn_a \frac{\partial a \left( \Delta_a \right)}{\partial \Delta_a} \, ,
\eea
 which is precisely \eqref{trial-c}.  

\section[Proof of the equality between \texorpdfstring{$c(b_i,\fn_a)$}{c(b(i),n(a))} and \texorpdfstring{$c_r(\Delta_a , \fn_a )$}{c[r](Delta(a),n(a))}]{Proof of the equality between $c(b_i,\fn_a)$ and $c_r(\Delta_a , \fn_a )$}
\label{proof}

In this appendix we prove \eqref{vector:u:constraint}.  As discussed in the text, \eqref{id1} and \eqref{id2} are simple consequences of this equation. 
We will also solve explicitly the equations \eqref{constraint} in a particular gauge. We first  review in appendix \ref{simplifications} some technical results of  \cite{Gauntlett:2018dpc} that will be used in the rest of the proof. For simplicity of notations, in this appendix $\Delta_a$ will always  refer to the quantities $\Delta_a(b_i,\fn_a)$ defined in \eqref{BZ}, unless otherwise stated. 

\subsection{Some simplifications}\label{simplifications}

The master volume is a quadratic form in $\lambda_a$,%
\footnote{To compare with \cite{Gauntlett:2018dpc}: $J_{ab}= (2\pi)^2 I_{ab}$.}
\bea
 \cV =\frac12 \sum_{a,b=1}^d J_{ab} \lambda_a \lambda_b \, ,
\eea
with a symmetric matrix $J_{ab}$, of rank $d-2$.  Indeed, it is invariant under
\bea
 \label{gauge}
 \lambda_a \rightarrow \lambda_a + \sum_{i=1}^3 l_i (b_1 v_a^i - b_i) \, ,
\eea
since, using \cite[(3.41)]{Gauntlett:2018dpc}, we find that
\bea
 \label{Id}
 \delta_{l_i} \cV = \sum_{a=1}^d \frac{\partial \cV}{ \partial \lambda_a} \delta_{l_i} \lambda_a = \sum_{a=1}^d \frac{\partial \cV}{\partial \lambda_a} (b_1 v_a^i - b_i) =0 \, .
\eea
Notice that this leaves  $d-2$ independent $\lambda_a$ since $l_1$ does not contribute ($v_a^1=1$ for all $a$). Correspondingly, the matrix $J_{a,a}$ has rank $d-2$.  
Since \eqref{Id} is valid for all $b_i$ and all $\lambda_a$ we obtain 
\bea\label{constr}  \sum_{a=1}^d \frac{\partial \cV}{\partial \lambda_a} (b_1 v_a^i - b_i)  = \sum_{a,b=1}^d J_{ab} \lambda_b (b_1 v_a^i - b_i) =0 \, \Longrightarrow  \sum_{a=1}^d  v_a^i J_{ab} = \frac{b_i}{b_1} \sum_{a=1}^d J_{ab} \, ,\eea
The equations \eqref{constraint} can be written as
\bea
 \label{eqs}  & N = -\sum_{a,b} J_{ab} \lambda_b \, , \\ 
 & \fn_a N = - \frac{A}{2\pi} \sum_{b} J_{ab} - b_1 \sum_{i,b} n^i \frac{\partial J_{ab}}{\partial b_i} \lambda_b \, , \\
 & A \sum_{ab} J_{ab} = 2\pi n^1 \sum_{ab} J_{ab} \lambda_b - 2\pi b_1 \sum_{a,b,i} n^i \frac{\partial J_{ab}}{\partial b_i} \lambda_b \, .
\eea
For given fluxes $\fn_a$ and number of colors $N$,  these are, in principle, $d+2$ equations for $d-1 =(d-2)+1$ variables $\lambda_a$ and $A$.
But, fortunately, three equations are redundant  \cite{Gauntlett:2018dpc}. Indeed the three linear combinations, $k=1,2,3$, of the equations for $\fn_a$ 
\bea
 \sum_a v_a^k \fn_a = -\frac{1}{2\pi N} \frac{b_k}{b_1} \bigg (  A \sum_{ab} J_{ab} - 2\pi n^1 \sum_{ab} J_{ab} \lambda_b + 2\pi b_1 \sum_{a,b,i} n^i \frac{\partial J_{ab}}{\partial b_i} \lambda_b \bigg ) - \frac{n^k}{N} \sum_{ab} J_{ab} \lambda_b \equiv n^k \, ,
\eea
reproduce the relation between $\fn_a$ and $n^k$. We used the first and third equations in \eqref{constraint}, the constraint \eqref{constr} and its derivative with respect to $b_i$ 
\bea
 \sum_a v_a^k  \frac{\partial J_{ab}}{\partial b_i} = \frac{b_k}{b_1} \sum_a  \frac{\partial J_{ab}}{\partial b_i}  +\left ( \delta^{ik} \frac{1}{b_1} - \delta^{i1}\frac{b_k}{(b_1)^2}\right ) \sum_a J_{ab} \, .
\eea 

We can also rewrite the functional  \eqref{Ssusy} as
\bea
 \label{cfunct}
 c(b_i,\fn_a) = - 48\pi^2 \bigg (A \sum_{a,b} J_{ab} \lambda_b +\pi b_1 \sum_{a,b,i} n^i  \frac{\partial J_{ab}}{\partial b_i} \lambda_a \lambda_b \bigg ) = 48 \pi^2 N \bigg (\frac{A }{2} + \pi \sum_{a=1}^d \lambda_a \fn_a \bigg )\, ,
\eea
where, in the second step,  we computed $\sum_a \lambda_a \fn_a$ from the second equation in \eqref{eqs}.

The R-charges \eqref{BZ} read
\bea
 \Delta_a = -\frac2N \frac{\partial {\cal V}}{\partial\lambda_a} = -\frac2N  \sum_b J_{ab} \lambda_b \, .
\eea
Multiplying \eqref{constr} by $-\frac2N \lambda_b$ and summing over $b$ we obtain 
\bea
 \label{Deltar0} \sum_{a=1}^d  v_a^i \Delta_a = -\frac2N   \frac{b_i}{b_1} \sum_{a,b=1}^d J_{ab} \lambda_b = 2 \frac{b_i}{b_1} \, .
\eea
Notice, in particular, that $\sum_a\Delta_a = 2$. Introducing the vectors $r_a=v_a - b/b_1$ we can also write
\bea
 \label{Deltar} \sum_{a=1}^d  r_a \Delta_a =0 \, ,
\eea  
an identity that we will use repeatedly in the following.  

\subsection{A convenient gauge}\label{sec:gauge}

We can simplify the equations choosing a gauge.  Using \eqref{gauge} we can set   two $\lambda_a$ to zero, say $\lambda_1=\lambda_2=0$. From \eqref{mastervol}, we see that the non-zero components of the matrix $J_{ab}$ are 
\bea
 J_{a,a+1} = 8 \pi^3 \frac{1}{ (v_a,v_{a+1},b)}\, , \qquad J_{a,a}= - 8 \pi^3 \frac{(v_{a-1},v_{a+1},b)}{ (v_{a-1},v_{a},b) (v_a,v_{a+1},b)} \, .
\eea
We also know that 
\bea
 \Delta_a = -\frac2N  \sum_{b=1}^d  J_{ab} \lambda_b \, .
\eea
In our gauge, $\lambda_1=\lambda_2=0$, we find that
\bea
 & \Delta_2 =  -\frac2N  (J_{23} \lambda_3)  \, , \\
 & \Delta_3=  -\frac2N  (J_{33} \lambda_3 +J_{34} \lambda_4 )  \, , \\
 & \Delta_4=  -\frac2N  (J_{43} \lambda_3+J_{44} \lambda_4 +J_{45} \lambda_5) \, ,
 & \qquad \ldots \, ,
\eea
that we can solve recursively. We obtain
\bea
 \label{eq1} \lambda_3 & =-\frac{N}{16\pi^3} (v_2,v_{3},b) \Delta_2  \, , \\
 \lambda_4 
 &= -\frac{N}{16\pi^3}( (v_2,v_{4},b) \Delta_2 + (v_3,v_{4},b) \Delta_3 ) \, , \\
 \lambda_5 & = -\frac{N}{16\pi^3} (v_4,v_{5},b) \left ( \Delta_4 + \frac{(v_3,v_{5},b)}{(v_4,v_{5},b)}  \Delta_3 + \frac{(v_2,v_{4},b)(v_3,v_{5},b) -(v_2,v_{3},b)(v_4,v_{5},b)}{(v_3,v_{4},b)(v_4,v_{5},b)} \Delta_2 \right ) \\
 & = -\frac{N}{16\pi^3}  \left ( (v_4,v_{5},b) \Delta_4 +(v_3,v_{5},b) \Delta_3 + (v_2,v_{5},b) \Delta_2 \right ) \, ,
\eea
where in the last step we used the identity
\be
 (A,B,b)(C,D,b) -(A,C,b)(B,D,b) - (A,D,b) (C,B,b)=0 \, ,
\ee
 valid for arbitrary vectors $A,B,C,D$. Altogether we can write  the inversion formula
\bea
 \label{inversion}
 \lambda_a = -\frac{N}{16\pi^3}  \sum_{c=2}^{a} (v_c,v_a,b) \Delta_c \, ,\qquad a=3,\ldots ,d \, .
\eea
As a consistency check, note that we can also extract $\lambda_d$ from the equation $\Delta_1 =-\frac2N J_{d1} \lambda_d$, obtaining
\bea
 \label{eq2}
 \lambda_d =  -\frac{N}{16\pi^3} (v_d,v_1,b) \Delta_1 \equiv  -\frac{N}{16\pi^3}  \sum_{c=2}^{d} (v_c,v_d,b) \Delta_c \, ,
\eea
where in the second step we used $\sum_c v_c \Delta_c =2 b/b_1$.
 
We can now analyze the equations \eqref{eqs} in the gauge $\lambda_1=\lambda_2=0$. Introduce the notation $\nabla \equiv \sum_i n^i \partial_{b_i}$. We have
\bea
 \nabla J_{a,a+1}= -8 \pi^3 \frac{(v_a,v_{a+1},n)}{(v_a,v_{a+1},b)^2} \, , \qquad \sum_b J_{ab}= 8\pi^3 b_1 \frac{(v_{a-1},v_a,v_{a+1})}{(v_{a-1}, v_a,b) (v_a,v_{a+1}, b)} \, ,
\eea
where, for the second identity, we used%
\footnote{\label{foot}The geometrical meaning of this identity is clear from figure \ref{toric:Reeb}: the sum of the areas of the plane triangles of vertices $(v_{a-1},v_a,B)$ and $(v_a,v_{a+1},B)$ minus the area of 
$(v_{a-1},v_{a+1},B)$ is the area of $(v_{a-1}, v_a,v_{a+1})$. The factor of $b_1$ takes into account the normalization in converting three-dimensional determinants into areas in the plane.}  
\bea
 \label{areas} (v_{a-1},v_a,b) +(v_a,v_{a+1},b)-(v_{a-1},v_{a+1},b) = b_1 (v_{a-1},v_a,v_{a+1}) \, .
\eea
Writing the equations \eqref{eqs} for $a=1$ and $a=2$
\bea
 & -N \fn_2 = \frac{A}{2\pi} (J_{12} +J_{22}+J_{23}) + b_1 \nabla J_{23} \lambda_3 \, ,\\
 & -N \fn_1 = \frac{A}{2\pi} (J_{1d} +J_{11}+J_{12}) + b_1 \nabla J_{1d} \lambda_d \, ,
\eea
and using \eqref{eq1} and \eqref{eq2} we find
\bea & \frac{N b_1}{16 \pi^3} (v_2,v_3,n) \Delta_2 = - \frac{A}{2\pi} b_1 \frac{(v_1,v_2,v_3)}{(v_1,v_2,b)} -\frac{(v_2,v_3,b)}{8\pi^3} N \fn_2 \, , \\
& \frac{N b_1}{16 \pi^3} (v_d,v_1,n) \Delta_1 = - \frac{A}{2\pi} b_1 \frac{(v_d,v_1,v_2)}{(v_1,v_2,b)}  - \frac{ (v_d ,v_1,b) }{ 8\pi^3 } N \fn_1 \, .  \eea
Multiplying the first by $(v_d,v_1,v_2)$ and the second by $(v_1,v_2,v_3)$ and subtracting we obtain
\bea \label{fundamentalidentity0}
&(v_d,v_1,v_2) (v_2,v_3,n) \Delta_2 - (v_1,v_2,v_3)(v_d,v_1,n) \Delta_1 \\ 
&= - \frac{2}{b_1} \left( (v_d,v_1,v_2)  (v_2,v_3,b) \fn_2 - (v_1,v_2,v_3)(v_d ,v_1,b) \fn_1\right ) \, .
\eea
This has been proved for $a=1$ and $a=2$ but should hold for all adjacent pairs $(a, a+1)$  because it is an identity for gauge invariant quantities and we can always
use an adapted gauge where $\lambda_a=\lambda_{a+1}=0$. Therefore, we find
\bea \label{fundamentalidentity}
&(v_{a-1},v_a,v_{a+1}) (v_{a+1},v_{a+2},n) \Delta_{a+1}- (v_a,v_{a+1},v_{a+2})(v_{a-1},v_{a},n) \Delta_a \\ 
&= - \frac{2}{b_1} \left( (v_{a-1},v_a,v_{a+1})  (v_{a+1},v_{a+2},b) \fn_{a+1} - (v_a,v_{a+1},v_{a+2})(v_{a-1} ,v_{a},b) \fn_a\right ) \, .
\eea
where we identify $\fn_{a+d} =\fn_a$, $\Delta_{a+d}=\Delta_a$ and $v_{a+d}=v_a$, so that for example $v_{d+1}=v_1$ and $v_0=v_d$. This is a set of equations that 
allow to find an explicit expression for  $\Delta_a$ using recursion to obtain $\Delta_{a+1}$ from $\Delta_a$ and enforcing $\sum_a \Delta_a=2$ in order to find the value of $\Delta_1$.
Notice that, at each step of the recursion, $b$ only appears linearly so  $\Delta_a$ is a regular function of $b_i$. $\Delta_a$ is actually a  linear polynomial  in $b/b_1$.

For further reference let us also quote the value of $A$:
\bea \label{area} A= -\frac{N}{8\pi^2} \frac{(v_2,v_3,n) (v_1,v_2,b) }{(v_1,v_2,v_3)}  \Delta_2 - \frac{N}{4\pi^2 b_1} \frac{(v_2,v_3,b) (v_1,v_2,b) }{(v_1,v_2,v_3)} \fn_2   \, .\eea
This equation and \eqref{inversion} guarantee that, in the gauge  $\lambda_1=\lambda_2=0$, $\lambda_a/b_1$ and $A/b_1$ are quadratic polynomials in $b/b_1$. 
  
\subsection{Completing the proof}
\label{proof:2}

We now prove  \eqref{vector:u:constraint} using the logic of  \cite{Butti:2005vn} and \cite{Lee:2006ru}.
In figure \ref{toric:Reeb} we draw the plane orthogonal to the vector $e_1=(1,0,0)$, where all the endpoints of the vectors $v_a$ lie.  The vectors $w_a=v_{a+1}-v_a$ lie entirely on the plane and correspond to the sides of the toric diagram. We also define the vectors $r_a = v_a - b/b_1$.  They also lie entirely on the plane and connect the point $B$  with coordinates $(b_2/b_1,b_3/b_2)$ to the vertices of the toric diagram. All the vectors in the following are three-dimensional.   We  use  $\langle C,D\rangle = (e_1,C,D)$ to compute areas in the plane. Indeed, when $C$ and $D$ are vectors lying on the plane, $|\langle C,D\rangle|$ is twice the area of the triangle with sides $C$ and $D$. We also assume that the vertices are labeled in the counterclockwise direction. 
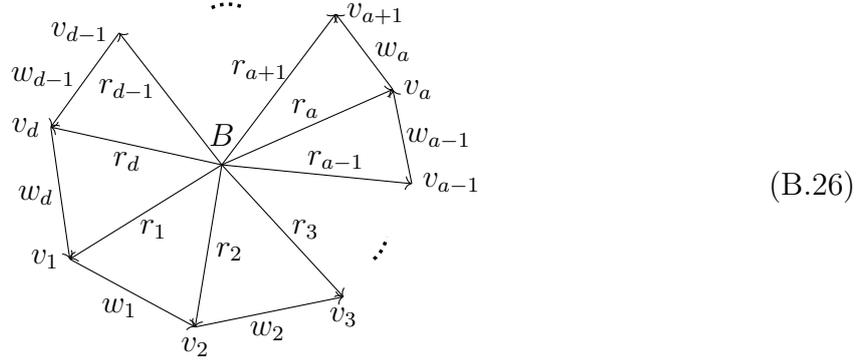
\bea
 \label{toric:Reeb}
 \begin{tikzpicture}
  [scale=0.5 ]
  \draw [line width=1.3, dotted] (4.,-2.5) arc [radius=1, start angle=-50, end angle=-10];
  
  \draw[->] (0,0) -- (5,-0.5) node[right] {$v_{a-1}$};
  \draw (3,0.1)  node {$r_{a-1}$};
  \draw[->] (5,-0.5) -- (4.5,2) node[left]{};
  \draw (5.7,0.7)  node {$w_{a-1}$};
  \draw[->] (0,0) -- (4.5,2) node[ right] {$v_a$};
  \draw (2.2,1.5)  node {$r_{a}$};
  \draw[->] (0,0) -- (3.,4) node[right] {$v_{a+1}$};
  \draw (1.,2.5)  node {$r_{a+1}$};
  \draw[->] (4.5,2) -- (3.,4.) node[left]{};
  \draw (4.5,3.)  node {$w_{a}$};
  
  \draw (0.,0.8)  node {$B$};
  \draw [line width=1.3, dotted] (0.5,4.2) arc [radius=1, start angle=70, end angle=120];
  
  \draw[->] (0,0) -- (-2.7,3.5) node[left] {$v_{d-1}$};
  \draw (-2.5,2.)  node {$r_{d-1}$};
  \draw[->] (-2.7,3.5) -- (-4.5,1.) node[right]{};
  \draw (-4.7,2.3)  node {$w_{d-1}$};
  \draw[->] (0,0) -- (-4.5,1.) node[left] {$v_{d}$};
  \draw (-2.5,0.1)  node {$r_{d}$};
  \draw[->] (0,0) -- (-4.,-2.5) node[left] {$v_{1}$};
  \draw (-1.8,-1.7)  node {$r_{1}$};
  \draw[->] (-4.5,1.) -- (-4.,-2.5) node[right]{};
  \draw (-4.9,-0.8)  node {$w_{d}$};
  \draw[->] (0,0) -- (-0.7,-4.3) node[below] {$v_{2}$};
  \draw (0.2,-2.3)  node {$r_{2}$};
  \draw[->] (-4.,-2.5) -- (-0.7,-4.3) node[right]{};
  \draw (-2.7,-3.8)  node {$w_{1}$};
  \draw[->] (-0.7,-4.3) -- (3.2,-3.5) node[right]{};
  \draw (1.2,-4.4)  node {$w_{2}$};
  \draw[->] (0,0) -- (3.2,-3.5) node[below] {$v_{3}$};
  \draw (2.2,-1.7)  node {$r_{3}$};
%  \draw[->] (3.2,-3.5) -- (5,-0.5) node[right]{};
%  \draw (4.6,-2.3)  node {$w_{3}$};
 \end{tikzpicture}
\eea

Define the quantities 
\bea
 \label{c_a:general}
 c_a = \sum_{b , c = 1}^{d} |(v_a,v_b,v_c)| \fn_b \Delta_c  = \sum_{b , c \in [a+1,a-1]} (v_a, v_b,v_c) (\fn_b \Delta_c + \fn_c \Delta_b) \, ,
\eea
where the symbol $b,c \in [A,B]$ means a sum over all pairs $A\le b < c\le B+d$ with the identification $\fn_{a+d} =\fn_a$, $\Delta_{a+d}=\Delta_a$ and $v_{a+d}=v_a$. If we select an order, we can drop the absolute value from $|(v_a,v_b,v_c)|$ but note that we need to keep the symmetrized product of $\fn$ and $\Delta$. We want to prove that   there exists a function $S$ and a vector $u$ such that
\bea
 \label{hj}
 c_a= S + \langle r_a, u \rangle \, ,
\eea
and $S$ is proportional to $c(b_i,\fn_a)$. Using repeatedly  the identity $(v_a,v_b,v_c)= \langle r_a,r_b \rangle +  \langle r_b,r_c \rangle -  \langle r_a,r_c \rangle$,%
\footnote{The geometrical interpretation of this identity is similar to the one discussed in  footnote \ref{foot}.}
we can compute the difference
\bea
 c_2 - c_1 & = \sum_{b , c \in [ 3 , d ]} (v_2 - v_1 , v_b , v_c) ( \fn_b \Delta_c + \fn_c \Delta_b )  + \sum_{b \in [ 3 , d ]} (v_2 , v_b , v_1) ( \fn_b \Delta_1 + \fn_1 \Delta_b ) \\ & - \sum_{c \in [ 3 , d ]} (v_1 , v_2 , v_c) ( \fn_2 \Delta_c + \fn_c \Delta_2 ) 
  = \sum_{b , c \in [ 3 , d ]} \langle w_1 , (r_b - r_c) ( \fn_b \Delta_c + \fn_c \Delta_b ) \rangle  \\
 & + \sum_{b \in [ 3 , d ]} \langle w_1 , (r_b - r_1) ( \fn_b \Delta_1 + \fn_1 \Delta_b ) \rangle 
  + \sum_{c \in [ 3 , d ]} \langle w_1 , (r_2 - r_c) ( \fn_2 \Delta_c + \fn_c \Delta_2 ) \rangle \\
 & = \sum_{b , c \in [ 2 , 1 ]} \langle w_1 , (r_b - r_c) ( \fn_b \Delta_c + \fn_c \Delta_b ) + d_1 w_1 \rangle
 \equiv \langle w_1 , u_1 \rangle\, ,
\eea
where we added an arbitrary term $d_1 w_1$ since it gives a vanishing contribution. More generally we find
\bea\label{diff} c_{a+1} - c_a = \langle  w_a , u_a \rangle \equiv \left\langle  w_a , \sum_{b,c \in [a+1,a]} (r_b -r_c) (\fn_b \Delta_c + \fn_c \Delta_b) + d_a w_a \right\rangle \, ,\eea
where again  $d_a$ is arbitrary. We will show now that there exists a choice of $d_a$ such that all the $u_a$ are equal. In order to show it, we compute the difference
\bea
 u_2 - u_1 & = \sum_{b , c \in [ 3 , 2 ]} (r_b - r_c) ( \fn_b \Delta_c + \fn_c \Delta_b ) - \sum_{b , c \in [ 2 , 1 ]} (r_b - r_c) ( \fn_b \Delta_c + \fn_c \Delta_b ) + d_2 w_2 - d_1 w_1 \\
 & = \sum_{a \in [3 , d]} [ (r_a - r_1) (\fn_a \Delta_1 + \fn_1 \Delta_a) + (r_a - r_2) (\fn_a \Delta_2 + \fn_2 \Delta_a)] + (r_1 - r_2) (\fn_1 \Delta_2 + \fn_2 \Delta_1)  \\
 & - \sum_{a \in [3 , d]} [ (r_a - r_1) (\fn_a \Delta_1 + \fn_1 \Delta_a) + (r_2 - r_a) (\fn_2 \Delta_a + \fn_a \Delta_2)] + (r_2 - r_1) (\fn_2 \Delta_1 + \fn_1 \Delta_2)  \\
 & + d_2 w_2 - d_1 w_1  = - 2 \sum_{a \in [3 , 1]} (r_2 - r_a) (\fn_a \Delta_2 + \fn_2 \Delta_a) + d_2 w_2 - d_1 w_1 \\
 & = - 2 \Big[ r_2 \Big ( \Delta_2 \sum_{a} \fn_a  +  2 \fn_2   \Big)  - \Delta_2 \big( \sum_{a} r_a \fn_a  \big)\Big ] + d_2 w_2 - d_1 w_1 \, ,
\eea
where we used $\sum_{b=1}^d \Delta_b=2$ and $\sum_{a=1}^d r_a \Delta_a =0$.
%, which follow from \eqref{Deltar}.  
More generally,
\bea\label{ppp} u_{a+1} - u_a = -2 \Big[ r_{a+1} \Big( \Delta_{a+1} \sum_b \fn_b + 2  \fn_{a+1}  \Big) - \Delta_{a+1} \sum_b r_b \fn_b \Big] + d_{a+1} w_{a+1} - d_a w_a \, .\eea 
%This expression should be zero and Mathematica confirms it. 
Requiring  that \eqref{ppp} is zero gives an expression for $d_a$.  Indeed, if $u_{a+1}=u_a$, contracting \eqref{ppp} with $w_{a}$ and $w_{a+1}$, respectively,  we obtain two different expressions for $d_a$:
\bea\label{pppp} &  d_{a+1} \langle w_a,w_{a+1} \rangle = 2 \Delta_{a+1} \left ( \langle w_a, r_{a+1}\rangle \sum_b \fn_b -\sum_b \fn_b \langle w_a, r_b\rangle   \right ) + 4 \fn_{a+1} \langle w_a, r_{a+1}\rangle  \, ,\\ 
& d_{a} \langle w_{a},w_{a+1} \rangle = 2 \Delta_{a+1} \left ( \langle w_{a+1}, r_{a+1}\rangle \sum_b \fn_b -\sum_b \fn_b \langle w_{a+1}, r_b\rangle   \right ) + 4 \fn_{a+1} \langle w_{a+1}, r_{a+1}\rangle  \, .
\eea
The two expressions should coincide. By shifting $a\rightarrow a+1$ in the second equation and comparing with the first one, we get the consistency condition 
\bea
 & \Delta_{a+2} \langle w_{a},w_{a+1}\rangle  \left ( \langle w_{a+2}, r_{a+2}\rangle \sum_b \fn_b -\sum_b \fn_b \langle w_{a+2}, r_b\rangle   \right )  +  2 \fn_{a+2} \langle w_{a},w_{a+1}\rangle  \langle w_{a+2}, r_{a+2}\rangle\\
 & - \Delta_{a+1}\langle w_{a+1},w_{a+2}\rangle  \left ( \langle w_a, r_{a+1}\rangle \sum_b \fn_b -\sum_b \fn_b \langle w_a, r_b\rangle   \right ) 
 -  2 \fn_{a+1} \langle w_{a+1},w_{a+2}\rangle \langle w_a, r_{a+1}  \rangle  =0 \, .
\eea
After a short computation  using  $\sum_a \fn_a v_a = n$ and  some geometrical identities to convert areas in the plane to three-dimensional determinants, like
\bea & \langle w_{a+2}, r_{a+2}\rangle \sum_b \fn_b -\sum_b \fn_b \langle w_{a+2}, r_b\rangle =    \langle w_{a+2},   n^1 v_{a+2} - n \rangle = -(v_{a+2},v_{a+3},n) \, , \\
& \langle w_a,w_{a+1} \rangle = (v_{a},v_{a+1},v_{a+2})\, ,  \qquad \langle w_{a}, r_{a}\rangle =-\frac1{b_1} (v_a,v_{a+1},b) \, ,
\eea
and generalizations, we obtain
\bea
 & \Delta_{a+2} (v_{a},v_{a+1},a_{a+2})  (v_{a+2},v_{a+3},n)  - \Delta_{a+1}  (v_{a+1},v_{a+2},v_{a+3})  (v_{a},v_{a+1},n)  = \\
 &- \frac{2}{b_1}\left ( \fn_{a+2} (v_{a},v_{a+1},v_{a+2}) (v_{a+2},v_{a+3},b) -\fn_{a+1} (v_{a+1},v_{a+2},v_{a+3}) (v_{a},v_{a+1},b)\right ) ,
\eea
which is precisely the identity \eqref{fundamentalidentity}. This  proves  that $d_a$ such that all the $u_a$ are equal can be actually found. 
We can also write an explicit expression  
\bea d_{a}  = - \frac{2  \Delta_{a+1} b_1 (v_{a+1},v_{a+2},n) + 4 \fn_{a+1} (v_{a+1},v_{a+2},b) } {b_1 (v_{a},v_{a+1},v_{a+2})}\, .\eea
Notice also that, comparing with \eqref{area}, we obtain
\bea
 \label{A:determined}
 A =  \frac{N}{16 \pi^2} (v_1,v_2,b) d_1 \, .
\eea

Since all the vectors $u_a$ are equal we call them $u$. We have
\bea c_{a+1}-c_a =\langle w_a, u \rangle  =\langle v_{a+1} - v_a, u \rangle   \Longrightarrow c_a= S + \langle r_a, u \rangle \, ,\eea
for some function $S$. Using \eqref{c_a:general} and \eqref{diff}, we find that
\bea
S = c_1 - \langle r_1, u \rangle 
%& = \sum_{b , c \in [2 , d]} (v_1 , v_b , v_c) ( \fn_b \Delta_c + \fn_c \Delta_b ) 
% - \left \langle r_1 , \sum_{b,c \in [2,1]} (r_b -r_c) (\fn_b \Delta_c + \fn_c \Delta_b) + d_1 w_1 \right \rangle \\
& = \sum_{b , c \in [2 , d]} \left( \langle r_1 , r_b \rangle + \langle r_b , r_c \rangle - \langle r_1 , r_c \rangle \right) ( \fn_b \Delta_c + \fn_c \Delta_b ) \\
& - \sum_{b,c \in [2,1]} \left( \langle r_1 , r_b \rangle - \langle r_1 , r_c \rangle \right) (\fn_b \Delta_c + \fn_c \Delta_b) - d_1 \langle r_1 , w_1 \rangle \\
%& = \sum_{b,c \in [2,d]}  \langle r_b , r_c \rangle ( \fn_b \Delta_c + \fn_c \Delta_b ) - \sum_{b\in [2,d]} \langle r_1 , r_b \rangle (\fn_b \Delta_1 + \fn_1 \Delta_b) - d_1 \langle r_1 , w_1 \rangle \, \\
%\eea
%Therefore, we find that
%\bea
%S & = \sum_{b,c\in [2,1]} \langle r_b,r_c \rangle (\fn_b \Delta_c + \fn_c \Delta_b)  - d_1 \langle r_1, w_1\rangle \\
& = \sum_{b,c\in [2,1]} \langle r_b,r_c \rangle (\fn_b \Delta_c + \fn_c \Delta_b) - \frac{16\pi^2  A}{b_1 N} \, .
\eea
 We can manipulate the sum in the previous expression by using repeatedly $\sum_c \Delta_c r_c=0$:
\bea & \sum_{b,c\in [2,1]} \langle r_b,r_c \rangle \fn_b \Delta_c +    \sum_{b,c\in [2,1]}  \langle r_b,r_c \rangle \fn_c \Delta_b = \sum_{c,b\in [2,1]} \langle r_c,r_b \rangle \fn_c \Delta_b +    \sum_{b,c\in [2,d]}  \langle r_b,r_c \rangle \fn_c \Delta_b \\
& = - \sum_{b,c\in [2,d]} \langle r_c,r_b \rangle \fn_c \Delta_b +    \sum_{b,c\in [2,d]}  \langle r_b,r_c \rangle \fn_c \Delta_b =  2 \sum_{b,c\in [2,d]}  \langle r_b,r_c \rangle \fn_c \Delta_b\, ,\eea
where in the first step we change variables in the first sum and notice that the term $c=1$ in the second sum is zero because of $\sum_c r_c \Delta_c =0$. In the second step, for each fixed $c$,
we transform  $\sum_{c<b\le 1} r_b \Delta_b = -\sum_{2\le b\le c} r_b \Delta_b$ and notice that $c\le d$.  In conclusion we have
 \bea S = 2 \sum_{b,c\in [2,d]}  \langle r_b,r_c \rangle \fn_c \Delta_b -\frac{16\pi^2 A}{b_1N } \, .
\eea
This has to be compared with \eqref{cfunct}:
\bea
 c_{r} 
  %= 48 \pi^2 N \left ( \frac{A}{2} + \pi \sum_{a=1}^d \lambda_a \fn_a \right ) 
  =  48 \pi^2 N \bigg ( \frac{A}{2} - \frac{N b_1}{16 \pi^2} \sum_{a=3}^d \sum_{c=2}^{a} \langle r_c,r_a\rangle \Delta_c \fn_a \bigg ) 
% & = 48 \pi^2 N \left ( \frac{A }{2} - \frac{N b_1}{16 \pi^2} \sum_{c , a \in [2,d]} \langle r_c, r_a\rangle \Delta_c \fn_a \right )\, , \\
  = 48 \pi^2 N \bigg ( \frac{A}{2} - \frac{N b_1}{16 \pi^2} \sum_{b , c \in [2,d]} \langle r_b, r_c\rangle \fn_c \Delta_b \bigg ) \, ,
\eea
where we worked in the gauge $\lambda_1=\lambda_2=0$ and we used \eqref{inversion} and the identity $(v_c,v_a,b)=b_1 \langle r_c,r_a \rangle $.
We see that
\bea
 S = -\frac{2}{3 N^2 b_1} c_{r} \, .
\eea
Multiplying \eqref{hj} by $-3 N^2$, and using the definition \eqref{c_a:general} we finally obtain
\bea 
-6 \sum_{b,c=1}^d c_{abc} \fn_b \Delta_c  = - 3 N^2 \sum_{b , c = 1}^{d} |(v_a,v_b,v_c)| \fn_b \Delta_c   = c_r + (e_1,r_a, t) \, ,\eea
where we set $b_1=2$ and $t= - 3 N^2 u$. This is exactly  \eqref{vector:u:constraint} .

\subsection[Proving that \texorpdfstring{$\cV_{\text{on-shell}} (b_i,\fn_a) = a (\Delta_a)$}{V[on-shell](b(i),n(a))=a(Delta(a))}]{Proving that $\mathcal{V}_{\text{on-shell}} (b_i,\mathfrak{n}_a) = a (\Delta_a)$}

In the gauge $\lambda_1=\lambda_2=0$, the master volume \eqref{mastervol} can be written as 
\bea
 \cV & = - \frac{N}{4} \sum_{c = 3}^d \lambda_c \Delta_c = \frac{N^2}{64 \pi^3} \sum_{c = 3}^d \sum_{b=2}^{c} (v_b ,v_c, b) \Delta_b \Delta_c 
  = \frac{N^2 b_1}{128 \pi^3} \sum_{c = 3}^d \sum_{b = 2}^{c} \sum_{a = 1}^{d} (v_b ,v_c, v_a) \Delta_a \Delta_b \Delta_c \\
 & = \frac{N^2 b_1}{128 \pi^3} \sum_{a = 1}^{d} \sum_{b , c \in [ 2 , d ]} (v_a ,v_b, v_c) \Delta_a \Delta_b \Delta_c
  = \frac{N^2 b_1}{128 \pi^3} \sum_{1\le a<b<c\le d} (v_a , v_b , v_c) \Delta_a \Delta_b \Delta_c
 \, ,
\eea
where we used \eqref{inversion} and $b = \frac{b_1}{2} \sum_a v_a \Delta_a$.%
\footnote{In the last step of the proof we reorganized  the sum  by noticing that the term  $ (v_1 ,v_e, v_f) \Delta_1 \Delta_e \Delta_f$ with $2\le  e <f \le d$
appears just once while  $(v_d ,v_e, v_f) \Delta_d \Delta_e \Delta_f$ with $2\le d < e <f \le d$ appears three times, two with the sign plus and one with sign minus.}
Comparing with \eqref{trial-a}, we see that
\be
 \cV_{\text{on-shell}} (b_i,\fn_a) =  \frac{b_1}{108 \pi^3 } a (\Delta_a) \Big |_{\Delta_a = \Delta_a(b,\fn)} \, .
\ee

\end{appendix}

\bibliographystyle{ytphys}

\bibliography{c-extremization}

\providecommand{\href}[2]{#2}\begingroup\raggedright\begin{thebibliography}{10}

\bibitem{Intriligator:2003jj}
K.~A. Intriligator and B.~Wecht, ``{The Exact superconformal R symmetry
  maximizes a},'' \href{http://dx.doi.org/10.1016/S0550-3213(03)00459-0}{{\em
  Nucl. Phys.} {\bfseries B667} (2003) 183--200},
\href{http://arxiv.org/abs/hep-th/0304128}{{\ttfamily arXiv:hep-th/0304128
  [hep-th]}}.
%%CITATION = HEP-TH/0304128;%%.

\bibitem{Benini:2012cz}
F.~Benini and N.~Bobev, ``{Exact two-dimensional superconformal R-symmetry and
  c-extremization},''
  \href{http://dx.doi.org/10.1103/PhysRevLett.110.061601}{{\em Phys. Rev.
  Lett.} {\bfseries 110} no.~6, (2013) 061601},
\href{http://arxiv.org/abs/1211.4030}{{\ttfamily arXiv:1211.4030 [hep-th]}}.
%%CITATION = ARXIV:1211.4030;%%.

\bibitem{Benini:2013cda}
F.~Benini and N.~Bobev, ``{Two-dimensional SCFTs from wrapped branes and
  c-extremization},'' \href{http://dx.doi.org/10.1007/JHEP06(2013)005}{{\em
  JHEP} {\bfseries 06} (2013) 005},
\href{http://arxiv.org/abs/1302.4451}{{\ttfamily arXiv:1302.4451 [hep-th]}}.
%%CITATION = ARXIV:1302.4451;%%.

\bibitem{Martelli:2005tp}
D.~Martelli, J.~Sparks, and S.-T. Yau, ``{The Geometric dual of a-maximisation
  for Toric Sasaki-Einstein manifolds},''
  \href{http://dx.doi.org/10.1007/s00220-006-0087-0}{{\em Commun. Math. Phys.}
  {\bfseries 268} (2006) 39--65},
\href{http://arxiv.org/abs/hep-th/0503183}{{\ttfamily arXiv:hep-th/0503183
  [hep-th]}}.
%%CITATION = HEP-TH/0503183;%%.

\bibitem{Martelli:2006yb}
D.~Martelli, J.~Sparks, and S.-T. Yau, ``{Sasaki-Einstein manifolds and volume
  minimisation},'' \href{http://dx.doi.org/10.1007/s00220-008-0479-4}{{\em
  Commun. Math. Phys.} {\bfseries 280} (2008) 611--673},
\href{http://arxiv.org/abs/hep-th/0603021}{{\ttfamily arXiv:hep-th/0603021
  [hep-th]}}.
%%CITATION = HEP-TH/0603021;%%.

\bibitem{Tachikawa:2005tq}
Y.~Tachikawa, ``{Five-dimensional supergravity dual of a-maximization},''
  \href{http://dx.doi.org/10.1016/j.nuclphysb.2005.11.010}{{\em Nucl. Phys.}
  {\bfseries B733} (2006) 188--203},
\href{http://arxiv.org/abs/hep-th/0507057}{{\ttfamily arXiv:hep-th/0507057
  [hep-th]}}.
%%CITATION = HEP-TH/0507057;%%.

\bibitem{Szepietowski:2012tb}
P.~Szepietowski, ``{Comments on a-maximization from gauged supergravity},''
  \href{http://dx.doi.org/10.1007/JHEP12(2012)018}{{\em JHEP} {\bfseries 12}
  (2012) 018},
\href{http://arxiv.org/abs/1209.3025}{{\ttfamily arXiv:1209.3025 [hep-th]}}.
%%CITATION = ARXIV:1209.3025;%%.

\bibitem{Karndumri:2013iqa}
P.~Karndumri and E.~O~Colgain, ``{Supergravity dual of $c$-extremization},''
  \href{http://dx.doi.org/10.1103/PhysRevD.87.101902}{{\em Phys. Rev.}
  {\bfseries D87} no.~10, (2013) 101902},
\href{http://arxiv.org/abs/1302.6532}{{\ttfamily arXiv:1302.6532 [hep-th]}}.
%%CITATION = ARXIV:1302.6532;%%.

\bibitem{Butti:2005vn}
A.~Butti and A.~Zaffaroni, ``{R-charges from toric diagrams and the equivalence
  of a-maximization and Z-minimization},''
  \href{http://dx.doi.org/10.1088/1126-6708/2005/11/019}{{\em JHEP} {\bfseries
  11} (2005) 019},
\href{http://arxiv.org/abs/hep-th/0506232}{{\ttfamily arXiv:hep-th/0506232
  [hep-th]}}.
%%CITATION = HEP-TH/0506232;%%.

\bibitem{Lee:2006ru}
S.~Lee and S.-J. Rey, ``{Comments on anomalies and charges of toric-quiver
  duals},'' \href{http://dx.doi.org/10.1088/1126-6708/2006/03/068}{{\em JHEP}
  {\bfseries 03} (2006) 068},
\href{http://arxiv.org/abs/hep-th/0601223}{{\ttfamily arXiv:hep-th/0601223
  [hep-th]}}.
%%CITATION = HEP-TH/0601223;%%.

\bibitem{Eager:2010yu}
R.~Eager, ``{Equivalence of A-Maximization and Volume Minimization},''
  \href{http://dx.doi.org/10.1007/JHEP01(2014)089}{{\em JHEP} {\bfseries 01}
  (2014) 089},
\href{http://arxiv.org/abs/1011.1809}{{\ttfamily arXiv:1011.1809 [hep-th]}}.
%%CITATION = ARXIV:1011.1809;%%.

\bibitem{Couzens:2018wnk}
C.~Couzens, J.~P. Gauntlett, D.~Martelli, and J.~Sparks, ``{A geometric dual of
  $c$-extremization},''
\href{http://arxiv.org/abs/1810.11026}{{\ttfamily arXiv:1810.11026 [hep-th]}}.
%%CITATION = ARXIV:1810.11026;%%.

\bibitem{Gauntlett:2018dpc}
J.~P. Gauntlett, D.~Martelli, and J.~Sparks, ``{Toric geometry and the dual of
  $c$-extremization},''
\href{http://arxiv.org/abs/1812.05597}{{\ttfamily arXiv:1812.05597 [hep-th]}}.
%%CITATION = ARXIV:1812.05597;%%.

\bibitem{Hanany:2005ve}
A.~Hanany and K.~D. Kennaway, ``{Dimer models and toric diagrams},''
\href{http://arxiv.org/abs/hep-th/0503149}{{\ttfamily arXiv:hep-th/0503149
  [hep-th]}}.
%%CITATION = HEP-TH/0503149;%%.

\bibitem{Franco:2005rj}
S.~Franco, A.~Hanany, K.~D. Kennaway, D.~Vegh, and B.~Wecht, ``{Brane dimers
  and quiver gauge theories},''
  \href{http://dx.doi.org/10.1088/1126-6708/2006/01/096}{{\em JHEP} {\bfseries
  01} (2006) 096},
\href{http://arxiv.org/abs/hep-th/0504110}{{\ttfamily arXiv:hep-th/0504110
  [hep-th]}}.
%%CITATION = HEP-TH/0504110;%%.

\bibitem{Feng:2005gw}
B.~Feng, Y.-H. He, K.~D. Kennaway, and C.~Vafa, ``{Dimer models from mirror
  symmetry and quivering amoebae},''
  \href{http://dx.doi.org/10.4310/ATMP.2008.v12.n3.a2}{{\em Adv. Theor. Math.
  Phys.} {\bfseries 12} no.~3, (2008) 489--545},
\href{http://arxiv.org/abs/hep-th/0511287}{{\ttfamily arXiv:hep-th/0511287
  [hep-th]}}.
%%CITATION = HEP-TH/0511287;%%.

\bibitem{Hosseini:2016cyf}
S.~M. Hosseini, A.~Nedelin, and A.~Zaffaroni, ``{The Cardy limit of the
  topologically twisted index and black strings in AdS$_{5}$},''
  \href{http://dx.doi.org/10.1007/JHEP04(2017)014}{{\em JHEP} {\bfseries 04}
  (2017) 014},
\href{http://arxiv.org/abs/1611.09374}{{\ttfamily arXiv:1611.09374 [hep-th]}}.
%%CITATION = ARXIV:1611.09374;%%.

\bibitem{Herzog:2010hf}
C.~P. Herzog, I.~R. Klebanov, S.~S. Pufu, and T.~Tesileanu, ``{Multi-Matrix
  Models and Tri-Sasaki Einstein Spaces},''
  \href{http://dx.doi.org/10.1103/PhysRevD.83.046001}{{\em Phys. Rev.}
  {\bfseries D83} (2011) 046001},
\href{http://arxiv.org/abs/1011.5487}{{\ttfamily arXiv:1011.5487 [hep-th]}}.
%%CITATION = ARXIV:1011.5487;%%.

\bibitem{Jafferis:2011zi}
D.~L. Jafferis, I.~R. Klebanov, S.~S. Pufu, and B.~R. Safdi, ``{Towards the
  F-Theorem: N=2 Field Theories on the Three-Sphere},''
  \href{http://dx.doi.org/10.1007/JHEP06(2011)102}{{\em JHEP} {\bfseries 06}
  (2011) 102},
\href{http://arxiv.org/abs/1103.1181}{{\ttfamily arXiv:1103.1181 [hep-th]}}.
%%CITATION = ARXIV:1103.1181;%%.

\bibitem{Hosseini:2016tor}
S.~M. Hosseini and A.~Zaffaroni, ``{Large $N$ matrix models for 3d ${\cal N}=2$
  theories: twisted index, free energy and black holes},''
  \href{http://dx.doi.org/10.1007/JHEP08(2016)064}{{\em JHEP} {\bfseries 08}
  (2016) 064},
\href{http://arxiv.org/abs/1604.03122}{{\ttfamily arXiv:1604.03122 [hep-th]}}.
%%CITATION = ARXIV:1604.03122;%%.

\bibitem{Benini:2015eyy}
F.~Benini, K.~Hristov, and A.~Zaffaroni, ``{Black hole microstates in AdS$_{4}$
  from supersymmetric localization},''
  \href{http://dx.doi.org/10.1007/JHEP05(2016)054}{{\em JHEP} {\bfseries 05}
  (2016) 054},
\href{http://arxiv.org/abs/1511.04085}{{\ttfamily arXiv:1511.04085 [hep-th]}}.
%%CITATION = ARXIV:1511.04085;%%.

\bibitem{Benini:2016rke}
F.~Benini, K.~Hristov, and A.~Zaffaroni, ``{Exact microstate counting for
  dyonic black holes in AdS4},''
  \href{http://dx.doi.org/10.1016/j.physletb.2017.05.076}{{\em Phys. Lett.}
  {\bfseries B771} (2017) 462--466},
\href{http://arxiv.org/abs/1608.07294}{{\ttfamily arXiv:1608.07294 [hep-th]}}.
%%CITATION = ARXIV:1608.07294;%%.

\bibitem{Hosseini:2016ume}
S.~M. Hosseini and N.~Mekareeya, ``{Large $N$ topologically twisted index:
  necklace quivers, dualities, and Sasaki-Einstein spaces},''
  \href{http://dx.doi.org/10.1007/JHEP08(2016)089}{{\em JHEP} {\bfseries 08}
  (2016) 089},
\href{http://arxiv.org/abs/1604.03397}{{\ttfamily arXiv:1604.03397 [hep-th]}}.
%%CITATION = ARXIV:1604.03397;%%.

\bibitem{Azzurli:2017kxo}
F.~Azzurli, N.~Bobev, P.~M. Crichigno, V.~S. Min, and A.~Zaffaroni, ``{A
  universal counting of black hole microstates in AdS$_{4}$},''
  \href{http://dx.doi.org/10.1007/JHEP02(2018)054}{{\em JHEP} {\bfseries 02}
  (2018) 054},
\href{http://arxiv.org/abs/1707.04257}{{\ttfamily arXiv:1707.04257 [hep-th]}}.
%%CITATION = ARXIV:1707.04257;%%.

\bibitem{Klebanov:1998hh}
I.~R. Klebanov and E.~Witten, ``{Superconformal field theory on three-branes at
  a Calabi-Yau singularity},''
  \href{http://dx.doi.org/10.1016/S0550-3213(98)00654-3}{{\em Nucl. Phys.}
  {\bfseries B536} (1998) 199--218},
\href{http://arxiv.org/abs/hep-th/9807080}{{\ttfamily arXiv:hep-th/9807080
  [hep-th]}}.
%%CITATION = HEP-TH/9807080;%%.

\bibitem{Acharya:1998db}
B.~S. Acharya, J.~M. Figueroa-O'Farrill, C.~M. Hull, and B.~J. Spence,
  ``{Branes at conical singularities and holography},''
  \href{http://dx.doi.org/10.4310/ATMP.1998.v2.n6.a2}{{\em Adv. Theor. Math.
  Phys.} {\bfseries 2} (1999) 1249--1286},
\href{http://arxiv.org/abs/hep-th/9808014}{{\ttfamily arXiv:hep-th/9808014
  [hep-th]}}.
%%CITATION = HEP-TH/9808014;%%.

\bibitem{Morrison:1998cs}
D.~R. Morrison and M.~R. Plesser, ``{Nonspherical horizons. 1.},''
  \href{http://dx.doi.org/10.4310/ATMP.1999.v3.n1.a1}{{\em Adv. Theor. Math.
  Phys.} {\bfseries 3} (1999) 1--81},
\href{http://arxiv.org/abs/hep-th/9810201}{{\ttfamily arXiv:hep-th/9810201
  [hep-th]}}.
%%CITATION = HEP-TH/9810201;%%.

\bibitem{Gauntlett:2004yd}
J.~P. Gauntlett, D.~Martelli, J.~Sparks, and D.~Waldram, ``{Sasaki-Einstein
  metrics on S**2 x S**3},''
  \href{http://dx.doi.org/10.4310/ATMP.2004.v8.n4.a3}{{\em Adv. Theor. Math.
  Phys.} {\bfseries 8} no.~4, (2004) 711--734},
\href{http://arxiv.org/abs/hep-th/0403002}{{\ttfamily arXiv:hep-th/0403002
  [hep-th]}}.
%%CITATION = HEP-TH/0403002;%%.

\bibitem{Gauntlett:2004hh}
J.~P. Gauntlett, D.~Martelli, J.~F. Sparks, and D.~Waldram, ``{A New infinite
  class of Sasaki-Einstein manifolds},''
  \href{http://dx.doi.org/10.4310/ATMP.2004.v8.n6.a3}{{\em Adv. Theor. Math.
  Phys.} {\bfseries 8} no.~6, (2004) 987--1000},
\href{http://arxiv.org/abs/hep-th/0403038}{{\ttfamily arXiv:hep-th/0403038
  [hep-th]}}.
%%CITATION = HEP-TH/0403038;%%.

\bibitem{Cvetic:2005ft}
M.~Cvetic, H.~Lu, D.~N. Page, and C.~N. Pope, ``{New Einstein-Sasaki spaces in
  five and higher dimensions},''
  \href{http://dx.doi.org/10.1103/PhysRevLett.95.071101}{{\em Phys. Rev. Lett.}
  {\bfseries 95} (2005) 071101},
\href{http://arxiv.org/abs/hep-th/0504225}{{\ttfamily arXiv:hep-th/0504225
  [hep-th]}}.
%%CITATION = HEP-TH/0504225;%%.

\bibitem{Benvenuti:2004dy}
S.~Benvenuti, S.~Franco, A.~Hanany, D.~Martelli, and J.~Sparks, ``{An Infinite
  family of superconformal quiver gauge theories with Sasaki-Einstein duals},''
  \href{http://dx.doi.org/10.1088/1126-6708/2005/06/064}{{\em JHEP} {\bfseries
  06} (2005) 064},
\href{http://arxiv.org/abs/hep-th/0411264}{{\ttfamily arXiv:hep-th/0411264
  [hep-th]}}.
%%CITATION = HEP-TH/0411264;%%.

\bibitem{Benvenuti:2005ja}
S.~Benvenuti and M.~Kruczenski, ``{From Sasaki-Einstein spaces to quivers via
  BPS geodesics: L**p,q|r},''
  \href{http://dx.doi.org/10.1088/1126-6708/2006/04/033}{{\em JHEP} {\bfseries
  04} (2006) 033},
\href{http://arxiv.org/abs/hep-th/0505206}{{\ttfamily arXiv:hep-th/0505206
  [hep-th]}}.
%%CITATION = HEP-TH/0505206;%%.

\bibitem{Butti:2005sw}
A.~Butti, D.~Forcella, and A.~Zaffaroni, ``{The Dual superconformal theory for
  L**pqr manifolds},''
  \href{http://dx.doi.org/10.1088/1126-6708/2005/09/018}{{\em JHEP} {\bfseries
  09} (2005) 018},
\href{http://arxiv.org/abs/hep-th/0505220}{{\ttfamily arXiv:hep-th/0505220
  [hep-th]}}.
%%CITATION = HEP-TH/0505220;%%.

\bibitem{Franco:2005sm}
S.~Franco, A.~Hanany, D.~Martelli, J.~Sparks, D.~Vegh, and B.~Wecht, ``{Gauge
  theories from toric geometry and brane tilings},''
  \href{http://dx.doi.org/10.1088/1126-6708/2006/01/128}{{\em JHEP} {\bfseries
  01} (2006) 128},
\href{http://arxiv.org/abs/hep-th/0505211}{{\ttfamily arXiv:hep-th/0505211
  [hep-th]}}.
%%CITATION = HEP-TH/0505211;%%.

\bibitem{Butti:2005ps}
A.~Butti and A.~Zaffaroni, ``{From toric geometry to quiver gauge theory: The
  Equivalence of a-maximization and Z-minimization},''
  \href{http://dx.doi.org/10.1002/prop.200510276}{{\em Fortsch. Phys.}
  {\bfseries 54} (2006) 309--316},
\href{http://arxiv.org/abs/hep-th/0512240}{{\ttfamily arXiv:hep-th/0512240
  [hep-th]}}.
%%CITATION = HEP-TH/0512240;%%.

\bibitem{Henningson:1998gx}
M.~Henningson and K.~Skenderis, ``{The Holographic Weyl anomaly},''
  \href{http://dx.doi.org/10.1088/1126-6708/1998/07/023}{{\em JHEP} {\bfseries
  07} (1998) 023},
\href{http://arxiv.org/abs/hep-th/9806087}{{\ttfamily arXiv:hep-th/9806087
  [hep-th]}}.
%%CITATION = HEP-TH/9806087;%%.

\bibitem{Benvenuti:2006xg}
S.~Benvenuti, L.~A. Pando~Zayas, and Y.~Tachikawa, ``{Triangle anomalies from
  Einstein manifolds},''
  \href{http://dx.doi.org/10.4310/ATMP.2006.v10.n3.a4}{{\em Adv. Theor. Math.
  Phys.} {\bfseries 10} no.~3, (2006) 395--432},
\href{http://arxiv.org/abs/hep-th/0601054}{{\ttfamily arXiv:hep-th/0601054
  [hep-th]}}.
%%CITATION = HEP-TH/0601054;%%.

\bibitem{Gubser:1998fp}
S.~S. Gubser and I.~R. Klebanov, ``{Baryons and domain walls in an N=1
  superconformal gauge theory},''
  \href{http://dx.doi.org/10.1103/PhysRevD.58.125025}{{\em Phys. Rev.}
  {\bfseries D58} (1998) 125025},
\href{http://arxiv.org/abs/hep-th/9808075}{{\ttfamily arXiv:hep-th/9808075
  [hep-th]}}.
%%CITATION = HEP-TH/9808075;%%.

\bibitem{Gubser:1998vd}
S.~S. Gubser, ``{Einstein manifolds and conformal field theories},''
  \href{http://dx.doi.org/10.1103/PhysRevD.59.025006}{{\em Phys. Rev.}
  {\bfseries D59} (1999) 025006},
\href{http://arxiv.org/abs/hep-th/9807164}{{\ttfamily arXiv:hep-th/9807164
  [hep-th]}}.
%%CITATION = HEP-TH/9807164;%%.

\bibitem{Amariti:2017iuz}
A.~Amariti, L.~Cassia, and S.~Penati, ``{c-extremization from toric
  geometry},'' \href{http://dx.doi.org/10.1016/j.nuclphysb.2018.01.025}{{\em
  Nucl. Phys.} {\bfseries B929} (2018) 137--170},
\href{http://arxiv.org/abs/1706.07752}{{\ttfamily arXiv:1706.07752 [hep-th]}}.
%%CITATION = ARXIV:1706.07752;%%.

\bibitem{Hosseini:2018uzp}
S.~M. Hosseini, I.~Yaakov, and A.~Zaffaroni, ``{Topologically twisted indices
  in five dimensions and holography},''
  \href{http://dx.doi.org/10.1007/JHEP11(2018)119}{{\em JHEP} {\bfseries 11}
  (2018) 119},
\href{http://arxiv.org/abs/1808.06626}{{\ttfamily arXiv:1808.06626 [hep-th]}}.
%%CITATION = ARXIV:1808.06626;%%.

\bibitem{Benini:2015bwz}
F.~Benini, N.~Bobev, and P.~M. Crichigno, ``{Two-dimensional SCFTs from
  D3-branes},'' \href{http://dx.doi.org/10.1007/JHEP07(2016)020}{{\em JHEP}
  {\bfseries 07} (2016) 020},
\href{http://arxiv.org/abs/1511.09462}{{\ttfamily arXiv:1511.09462 [hep-th]}}.
%%CITATION = ARXIV:1511.09462;%%.

\bibitem{Amariti:2016mnz}
A.~Amariti and C.~Toldo, ``{Betti multiplets, flows across dimensions and
  c-extremization},'' \href{http://dx.doi.org/10.1007/JHEP07(2017)040}{{\em
  JHEP} {\bfseries 07} (2017) 040},
\href{http://arxiv.org/abs/1610.08858}{{\ttfamily arXiv:1610.08858 [hep-th]}}.
%%CITATION = ARXIV:1610.08858;%%.

\bibitem{Amariti:2017cyd}
A.~Amariti, L.~Cassia, and S.~Penati, ``{Surveying 4d SCFTs twisted on Riemann
  surfaces},'' \href{http://dx.doi.org/10.1007/JHEP06(2017)056}{{\em JHEP}
  {\bfseries 06} (2017) 056},
\href{http://arxiv.org/abs/1703.08201}{{\ttfamily arXiv:1703.08201 [hep-th]}}.
%%CITATION = ARXIV:1703.08201;%%.

\bibitem{Couzens:2017nnr}
C.~Couzens, D.~Martelli, and S.~Schafer-Nameki, ``{F-theory and
  AdS$_{3}$/CFT$_{2}$ (2, 0)},''
  \href{http://dx.doi.org/10.1007/JHEP06(2018)008}{{\em JHEP} {\bfseries 06}
  (2018) 008},
\href{http://arxiv.org/abs/1712.07631}{{\ttfamily arXiv:1712.07631 [hep-th]}}.
%%CITATION = ARXIV:1712.07631;%%.

\bibitem{Martelli:2004wu}
D.~Martelli and J.~Sparks, ``{Toric geometry, Sasaki-Einstein manifolds and a
  new infinite class of AdS/CFT duals},''
  \href{http://dx.doi.org/10.1007/s00220-005-1425-3}{{\em Commun. Math. Phys.}
  {\bfseries 262} (2006) 51--89},
\href{http://arxiv.org/abs/hep-th/0411238}{{\ttfamily arXiv:hep-th/0411238
  [hep-th]}}.
%%CITATION = HEP-TH/0411238;%%.

\bibitem{Benini:2015noa}
F.~Benini and A.~Zaffaroni, ``{A topologically twisted index for
  three-dimensional supersymmetric theories},''
  \href{http://dx.doi.org/10.1007/JHEP07(2015)127}{{\em JHEP} {\bfseries 07}
  (2015) 127},
\href{http://arxiv.org/abs/1504.03698}{{\ttfamily arXiv:1504.03698 [hep-th]}}.
%%CITATION = ARXIV:1504.03698;%%.

\bibitem{Benini:2016hjo}
F.~Benini and A.~Zaffaroni, ``{Supersymmetric partition functions on Riemann
  surfaces},'' {\em Proc. Symp. Pure Math.} {\bfseries 96} (2017) 13--46,
\href{http://arxiv.org/abs/1605.06120}{{\ttfamily arXiv:1605.06120 [hep-th]}}.
%%CITATION = ARXIV:1605.06120;%%.

\bibitem{Closset:2016arn}
C.~Closset and H.~Kim, ``{Comments on twisted indices in 3d supersymmetric
  gauge theories},'' \href{http://dx.doi.org/10.1007/JHEP08(2016)059}{{\em
  JHEP} {\bfseries 08} (2016) 059},
\href{http://arxiv.org/abs/1605.06531}{{\ttfamily arXiv:1605.06531 [hep-th]}}.
%%CITATION = ARXIV:1605.06531;%%.

\bibitem{Aharony:2008ug}
O.~Aharony, O.~Bergman, D.~L. Jafferis, and J.~Maldacena, ``{N=6 superconformal
  Chern-Simons-matter theories, M2-branes and their gravity duals},''
  \href{http://dx.doi.org/10.1088/1126-6708/2008/10/091}{{\em JHEP} {\bfseries
  10} (2008) 091},
\href{http://arxiv.org/abs/0806.1218}{{\ttfamily arXiv:0806.1218 [hep-th]}}.
%%CITATION = ARXIV:0806.1218;%%.

\bibitem{Cacciatori:2009iz}
S.~L. Cacciatori and D.~Klemm, ``{Supersymmetric AdS(4) black holes and
  attractors},'' \href{http://dx.doi.org/10.1007/JHEP01(2010)085}{{\em JHEP}
  {\bfseries 01} (2010) 085},
\href{http://arxiv.org/abs/0911.4926}{{\ttfamily arXiv:0911.4926 [hep-th]}}.
%%CITATION = ARXIV:0911.4926;%%.

\bibitem{DallAgata:2010ejj}
G.~Dall'Agata and A.~Gnecchi, ``{Flow equations and attractors for black holes
  in N = 2 U(1) gauged supergravity},''
  \href{http://dx.doi.org/10.1007/JHEP03(2011)037}{{\em JHEP} {\bfseries 03}
  (2011) 037},
\href{http://arxiv.org/abs/1012.3756}{{\ttfamily arXiv:1012.3756 [hep-th]}}.
%%CITATION = ARXIV:1012.3756;%%.

\bibitem{Hristov:2010ri}
K.~Hristov and S.~Vandoren, ``{Static supersymmetric black holes in AdS4 with
  spherical symmetry},'' \href{http://dx.doi.org/10.1007/JHEP04(2011)047}{{\em
  JHEP} {\bfseries 04} (2011) 047},
\href{http://arxiv.org/abs/1012.4314}{{\ttfamily arXiv:1012.4314 [hep-th]}}.
%%CITATION = ARXIV:1012.4314;%%.

\bibitem{Hanany:2008fj}
A.~Hanany, D.~Vegh, and A.~Zaffaroni, ``{Brane Tilings and M2 Branes},''
  \href{http://dx.doi.org/10.1088/1126-6708/2009/03/012}{{\em JHEP} {\bfseries
  03} (2009) 012},
\href{http://arxiv.org/abs/0809.1440}{{\ttfamily arXiv:0809.1440 [hep-th]}}.
%%CITATION = ARXIV:0809.1440;%%.

\bibitem{Alday:2009qq}
L.~F. Alday, F.~Benini, and Y.~Tachikawa, ``{Liouville/Toda central charges
  from M5-branes},''
  \href{http://dx.doi.org/10.1103/PhysRevLett.105.141601}{{\em Phys. Rev.
  Lett.} {\bfseries 105} (2010) 141601},
\href{http://arxiv.org/abs/0909.4776}{{\ttfamily arXiv:0909.4776 [hep-th]}}.
%%CITATION = ARXIV:0909.4776;%%.

\bibitem{Bah:2011vv}
I.~Bah, C.~Beem, N.~Bobev, and B.~Wecht, ``{AdS/CFT Dual Pairs from M5-Branes
  on Riemann Surfaces},''
  \href{http://dx.doi.org/10.1103/PhysRevD.85.121901}{{\em Phys. Rev.}
  {\bfseries D85} (2012) 121901},
\href{http://arxiv.org/abs/1112.5487}{{\ttfamily arXiv:1112.5487 [hep-th]}}.
%%CITATION = ARXIV:1112.5487;%%.

\bibitem{Bah:2012dg}
I.~Bah, C.~Beem, N.~Bobev, and B.~Wecht, ``{Four-Dimensional SCFTs from
  M5-Branes},'' \href{http://dx.doi.org/10.1007/JHEP06(2012)005}{{\em JHEP}
  {\bfseries 06} (2012) 005},
\href{http://arxiv.org/abs/1203.0303}{{\ttfamily arXiv:1203.0303 [hep-th]}}.
%%CITATION = ARXIV:1203.0303;%%.

\end{thebibliography}\endgroup

\end{document}